\documentclass[modern,trackchanges]{aastex62}
\usepackage{CJKutf8}

\newcommand{\dml}{dm-$\lambda$}
\newcommand{\cntext}[1]{\begin{CJK}{UTF8}{gbsn}#1\end{CJK}}
\turnoffedit

\received{Nov 7, 2018}
\revised{Jan 13, 2019}
\accepted{Jan 15, 2019}

\submitjournal{ApJ}

\shorttitle{Subsecond-period Propagating Waves in a Flare}
\shortauthors{Yu \& Chen}

\begin{document}
\title{Possible Detection of Subsecond-period Propagating Magnetohydrodynamics Waves in Post-reconnection Magnetic Loops during a Two-ribbon Solar Flare}

\correspondingauthor{Bin Chen}
\email{bin.chen@njit.edu}
\author[0000-0003-2872-2614]{Sijie Yu (\cntext{余思捷})}
\affil{Center for Solar-Terrestrial Research, New Jersey Institute of Technology,
323 Martin Luther King Jr. Blvd, Newark, NJ 07102-1982, USA}

\author[0000-0002-0660-3350]{Bin Chen (\cntext{陈彬})}
\affil{Center for Solar-Terrestrial Research, New Jersey Institute of Technology,
323 Martin Luther King Jr. Blvd, Newark, NJ 07102-1982, USA}

\begin{abstract}

Solar flares involve the sudden release of magnetic energy in the solar corona. Accelerated nonthermal electrons have often been invoked as the primary means for transporting the bulk of the released energy to the lower solar atmosphere. However, significant challenges remain for this scenario, especially in accounting for the large number of accelerated electrons inferred from observations. Propagating magnetohydrodynamics (MHD) waves, particularly those with subsecond/second-scale periods, have been proposed as an alternative means for transporting the released flare energy, \edit1{likely alongside the electron beams}, while observational evidence remains elusive. Here we report a possible detection of such waves in the \edit1{late} impulsive phase of a two-ribbon flare. This is based on ultrahigh cadence dynamic imaging spectroscopic observations of a peculiar type of decimetric radio bursts obtained by the Karl G. Jansky Very Large Array. Radio imaging at each time and frequency pixel allows us to trace the spatiotemporal motion of the source, which agrees with the implications of the frequency drift pattern in the dynamic spectrum. The radio source\edit1{, propagating at 1000--2000 km s$^{-1}$ in projection, shows close spatial and temporal association with transient brightenings on the flare ribbon}. In addition, multitudes of subsecond-period oscillations are present in the radio emission. We interpret the observed radio bursts as \edit1{short-period} MHD wave packets propagating along newly reconnected magnetic flux tubes linking to the flare ribbon. The estimated energy flux carried by the waves is comparable to that needed to account for the plasma heating \edit1{during the late impulsive phase of this flare}.
\end{abstract}


\section{Introduction} \label{sec:intro}
An outstanding question in solar flare studies is how a large amount of magnetic energy released in a flare (up to 10$^{33}$ erg) is converted into other forms of energy in accelerated particles, heated plasma, waves/turbulence, and bulk motions, and transported throughout the flare region. The collisional thick-target model (CTTM; \citealt{1971SoPh...18..489B}), along with the framework of the standard CSHKP flare scenario \citep{1964NASSP..50..451C, 1966Natur.211..695S, 1974SoPh...34..323H, 1976SoPh...50...85K}, assumes that \edit1{a considerable fraction} of the magnetic energy released via reconnection goes into acceleration of charged electrons and ions to nonthermal energies in the solar corona \edit1{\citep{2004JGRA..10910104E,2005JGRA..11011103E,2012ApJ...759...71E}}. The downward-propagating electrons along the reconnected, close field lines slam into the dense chromosphere and lose most of their energy through Coulomb collisions. This sudden energy loss results in the \edit1{intense} heating of the chromospheric material within a confined region at the footpoints of the closed arcades, driving hot and dense material upward and filling the arcades --- a process known as ``chromospheric evaporation.'' The arcades, in turn, accumulate a large emission measure at high temperatures, thereby appearing particularly bright in extreme ultraviolet (EUV) and soft X-ray (SXR) wavelengths (see, e.g., a recent review by \citealt{2017LRSP...14....2B}). 

The CTTM model has been successful in accounting for a variety of flare phenomena, most notably the ``Neupert effect'': The high-energy, hard X-ray (HXR) emission tends to coincide temporally with the rate of the rising lower-energy, SXR emission during the primary phase of energy release (also known as the ``impulsive phase'') of a flare \citep{1968ApJ...153L..59N,2002A&A...392..699V}. Other outstanding examples include the decreasing height \citep[e.g.][]{2002SoPh..210..383A,2002SoPh..210..373B,2011A&A...533L...2B,2016ApJ...818...44R} and area \citep{2008A&A...489L..57K} of HXR footpoint sources with increasing energy. However, significant challenges remain for the CTTM model (see, e.g., \citealt{2009A&A...508..993B} and references therein). One challenge is the so-called ``number problem'': the total number of nonthermal electrons required to account for the observed HXR, (E)UV, or white light (WL) footpoint sources or flare ribbons can be very large compared to that available in the corona \citep[e.g.,][]{2007ApJ...656.1187F,2011ApJ...739...96K}. Similar implications have been argued based on observations of coronal HXR sources --- the inferred number density of nonthermal electrons is a large fraction of, or in some cases, nearly equal to, the total electron density available in the corona \citep{2007ApJ...669L..49K,2008ApJ...678L..63K,2010ApJ...714.1108K}. \edit1{This requires electrons to replenished the corona at the same rate as nonthermal electrons precipitate from it, otherwise the coronal acceleration region would be quickly evacuated. A scenario that invokes return currents, which involve electrons flowing up from the chromosphere into the corona to neutralize the depletion of the coronal electrons, has been suggested to alleviate the difficulty \citep[see, e.g.,][]{1970PhFl...13.1831H,1976SoPh...48..197H,1977ApJ...218..306K,1980ApJ...235.1055E,1990A&A...234..496V,2012ApJ...745...52H,2017ApJ...851...78A}}. \edit1{Nevertheless, these considerations have led various authors to suggest alternative scenarios that invoke electron (re)acceleration in the lower, denser solar atmosphere \citep{2008ApJ...675.1645F,2009A&A...508..993B,2014A&A...563A..51V}.}  Other mechanisms have also been proposed for heating the chromospheric plasma, such as thermal conduction or magnetohydrodynamics (MHD) waves  \citep{2009A&A...498..891B,2016ApJ...818...44R,2016ApJ...818L..20R}. In all cases, alternative means, \edit1{possibly operating alongside accelerated electrons} as in the CTTM model, are postulated to transport a sizable portion of the released flare energy from the reconnection region, presumably located in the corona, downward to spatially confined regions in the lower solar atmosphere.

One excellent way to provide such focused energy transport other than electron beams is via propagating plasma waves within reconnected flare arcades \citep{1982SoPh...80...99E,2008ApJ...675.1645F,2013ApJ...765...81R,2013A&A...558A..76R}. A variety of plasma waves, including Alfv\'en waves and fast-mode and slow-mode magnetosonic waves, can arise as a natural consequence of the flare energy being released in an impulsive fashion (see, e.g., recent studies by \citealt{2017ApJ...847....1T} and \citealt{2018ApJ...860..138P}). As argued by \citet{2008ApJ...675.1645F} and \citet{2013ApJ...765...81R}, plasma waves are capable of carrying a significant amount of flare energy, which  may be comparable to that needed to power the radiative emissions of a flare. \edit1{An intriguing recent numerical study by \citet{2016ApJ...818L..20R} demonstrated that the waves can drive chromospheric evaporation in a strikingly similar fashion to the way electron beams do. Their results were then confirmed by \citet{2016ApJ...827..101K}, who further showed that the detailed shapes of certain chromospheric lines could be used as a potential observational test to distinguish between the wave- and electron-beam heating scenarios.} 

Observationally, flare-associated quasi-periodic pulsations (QPPs) with different periods ranging from $<$1 s to tens of minutes have been detected at virtually all wavelengths. One of the main origins for the QPPs is thought to be MHD oscillations or waves (see, e.g., \citealt{2009SSRv..149..119N} for a review). Observational evidence for large-scale wave-like phenomena associated with flares has also frequently been reported using spatially-resolved imaging data (see reviews by e.g., \citealt{2012SoPh..281..187P, 2014SoPh..289.3233L, 2015LRSP...12....3W, 2017SoPh..292....7L}, a study of a large sample of such events in \citealt{2013ApJ...776...58N}, and a most recent observation of the 2017 September 10 X8.2 flare in \citealt{2018ApJ...864L..24L}). Observational evidence that links the response in the lower solar atmosphere to downward-propagating MHD waves, however, is rather rare. One outstanding example was from \citet{2016NatCo...713104L}, who found a sudden sunspot rotation during the impulsive phase of a flare based on observations from the Goode Solar Telescope of the Big Bear Solar Observatory (GST/BBSO), possibly triggered by downward-propagating waves generated by the release of flare energy. Another interesting study by \citet{2015ApJ...810....4B} reported long-period ($\sim$140 s), slow ($\sim$20 km s$^{-1}$) oscillating flare ribbons based on observations by the \textit{Interface Region Imaging Spectrograph}, although the authors interpreted the oscillating phenomenon in terms of instabilities in the reconnection current sheet rather than MHD waves. It is worthwhile to point out that, in the Earth's magnetosphere, direct evidence for Alfv\'en waves propagating along the outer boundary of the ``plasma sheet'' (which is analogous to newly reconnected flare loops) has been reported based on \textit{in situ} measurements. These waves have been argued to be responsible in transporting a significant amount of energy flux (in the form of Poynting flux) from the site of energy release in the magnetotail toward the Earth, which, in turn, powers the auroral emission that is analogous to flare ribbons on the Sun \citep{2000JGR...10518675W,2002JGRA..107.1201W,2000GeoRL..27.3169K}. 

Recently, numerical and analytical models have been developed to investigate \edit1{energy transport and deposition} from the corona to the low solar atmosphere by MHD waves \citep{2013ApJ...765...81R,2016ApJ...827..101K,2016ApJ...818L..20R,2017ApJ...847....1T,2018ApJ...853..101R}. An important finding is that short-period MHD waves, especially those having periods of about one second or less, carry a significant amount of energy \citep{2017ApJ...847....1T}, suffer much less energy loss when propagating out from the corona to the lower solar atmosphere \citep{2013ApJ...765...81R,2018ApJ...860..138P}, and are much more efficient in dissipating the energy in the upper chromosphere than their long-period counterparts \citep{2008ApJ...675.1645F,2013ApJ...765...81R,2016ApJ...818L..20R}. Therefore, these short-period MHD waves are thought to be a potential candidate for an alternative carrier for energy released in flares. Subsecond-period ($P<1$ s) QPPs have frequently been reported in radio and X-ray light curves and/or dynamic spectra (e.g., \citealt{1970A&A.....9..159R,1983SvAL....9..163B,2007SoPh..246..431C,2010ApJ...723...25T,2013ApJ...777..159Y}). \edit1{However, most of the large-scale wave-like phenomena detected on the basis of imaging data fall into the long-period regime ($>$10 s, e.g., \citealt{2009SSRv..149..119N}), with some rare exceptions from eclipse observations \citep[e.g.,][]{2002SoPh..207..241P}.} This is mainly due to the limitation on temporal cadence of current WL/EUV imaging instrumentation, or the lack of radio/X-ray imaging capability at high temporal cadence with sufficient dynamic range or counting statistics.

Here we report ultrahigh cadence (0.05 s) spectroscopic imaging of a peculiar type of radio bursts in the decimetric wavelength range (``\dml'' hereafter) that is likely associated with subsecond-period MHD waves propagating along flaring arcades. The bursts were recorded by the Karl G. Jansky Very Large Array (VLA) in a \textit{GOES}-class C7.2 flare that is associated with a failed filament eruption and large-scale coronal EUV waves. We further show that these MHD waves may carry a significant amount of energy flux that is comparable to the average energy flux needed for driving the plasma heating at the flare ribbons. In Section \ref{sec-obs}, we present VLA dynamic imaging spectroscopic observations of the radio bursts, supported by complementary magnetic, EUV, and X-ray data. In Section \ref{sec-discussion}, we interpret the observations within a physical scenario that involves propagating short-period MHD wave packets and discuss their energetics. We briefly summarize our findings in Section \ref{sec-conclusion}.

\section{Observations}\label{sec-obs}

\subsection{Event overview}\label{sec-overview}
The VLA is a general-purpose radio interferometer operating at $<$1--50 GHz. It has completed  a major upgrade \citep{2011ApJ...739L...1P} and was partially commissioned for solar observation in late 2011 \citep{2013ApJ...763L..21C}. It is capable of making broadband radio imaging spectroscopic observations in more than one thousand spectral channels with ultrahigh time resolution of tens of millisecond-scale.  Recent studies with the VLA have demonstrated its unique power in using coherent solar radio bursts to diagnose the production and transport of energetic electrons in solar flares by utilizing its imaging capabilities with spectrometer-like time and spectral resolution \citep{2013ApJ...763L..21C,2014ApJ...794..149C, 2015Sci...350.1238C,2018ApJ...866...62C,2017ApJ...848...77W}. 

\begin{figure*}[ht]
\includegraphics[width=\textwidth]{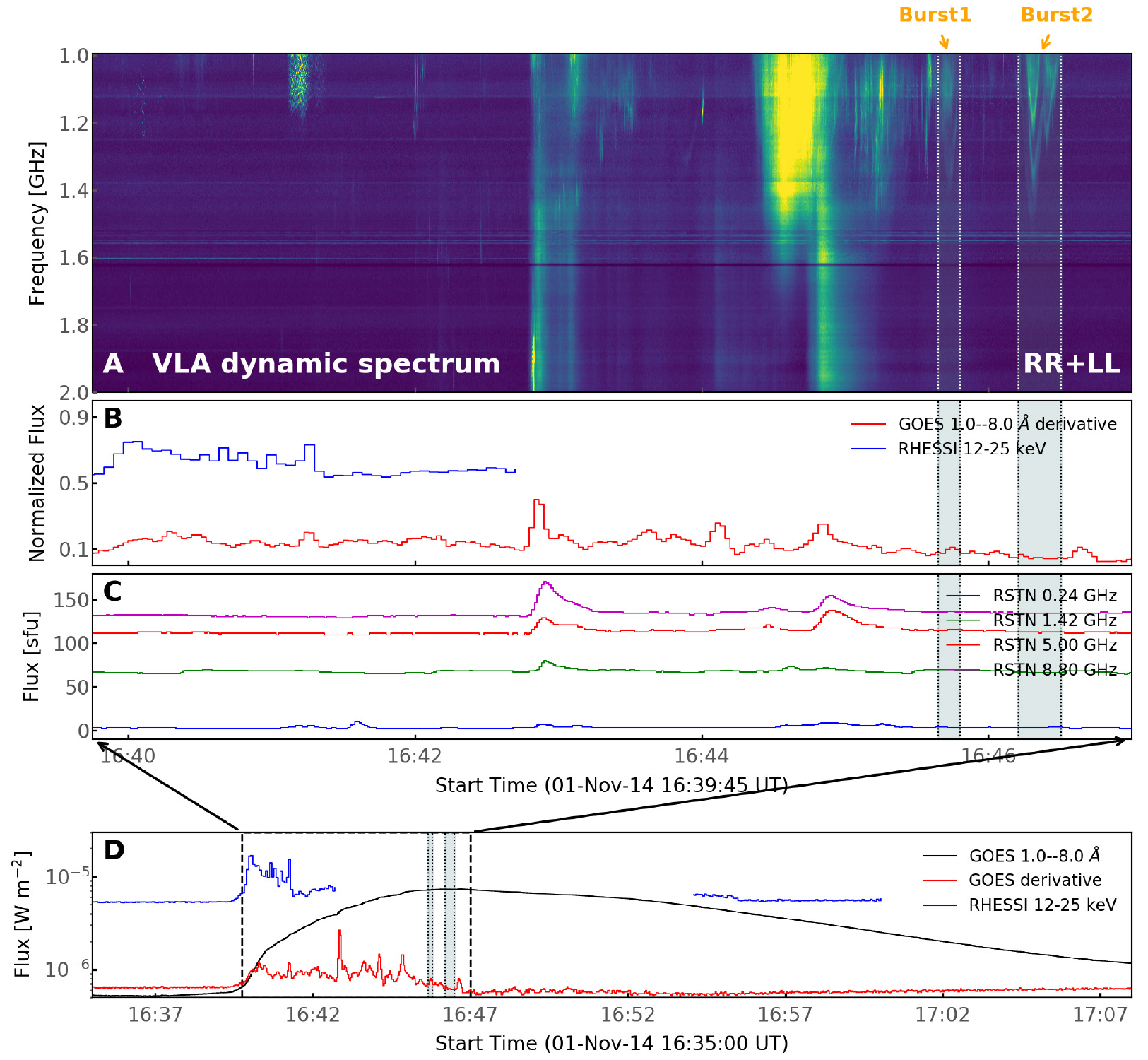}
\caption{(A) VLA cross-power dynamic spectrum at 1-2 GHz of the impulsive phase of the SOL2014-11-01T16:39 event. The frequency axis is inverted with higher frequency shown at the bottom. (B) \textit{RHESSI} 12--25 keV light curve (blue) and the time derivative of the \textit{GOES} 1--8 \AA\ soft X-ray light curves (red). The periods of the two radio bursts are bracketed by vertical dashed lines. (C) RSTN radio flux at multiple frequencies from 0.24 to 8.8 GHz. (D) \textit{GOES} and \textit{RHESSI} light curves of the entire event.  \label{fig-overview}}
\end{figure*}

The event under investigation occurred on 2014 November 11 in NOAA active region (AR) $12201$, located at $44\degr$ east from the central meridian. It is a \textit{GOES}-class C7.2 solar flare (flare identifier ``SOL2014-11-01T16:39:00L085C095'' following the IAU convention suggested by \citealt{2010SoPh..263....1L}). This event was well observed by the Atmospheric Imaging Assembly (AIA; \citealt{2012SoPh..275...17L}) and the Helioseismic and Magnetic Imager (HMI; \citealt{2012SoPh..275..207S}) aboard the \textit{Solar Dynamics Observatory} (\textit{SDO}). The impulsive phase of the flare started from $\sim$16:39 UT and was partially covered by \textit{RHESSI} \citep{2002SoPh..210....3L} until 16:42 UT, when the spacecraft entered the South Atlantic Anomaly (SAA). The VLA was used to observe the Sun from 16:30:10 UT to 20:40:09 UT and captured the entire flare. The observations were made in frequency bands between 1 and 2 GHz with 50 ms cadence and 2 MHz spectral resolution in dual circular polarizations. The 27-antenna array was in the C configuration (maximum baseline length 3 km), yielding an intrinsic angular resolution of $35''.7\times 16''.3$ at $\nu=1$ GHz at the time of the observation (and this scales linearly with $1/\nu$). The deconvolved synthesis images are restored with a $30''$ circular beam. 

\begin{figure*}[ht]
\includegraphics[width=\textwidth]{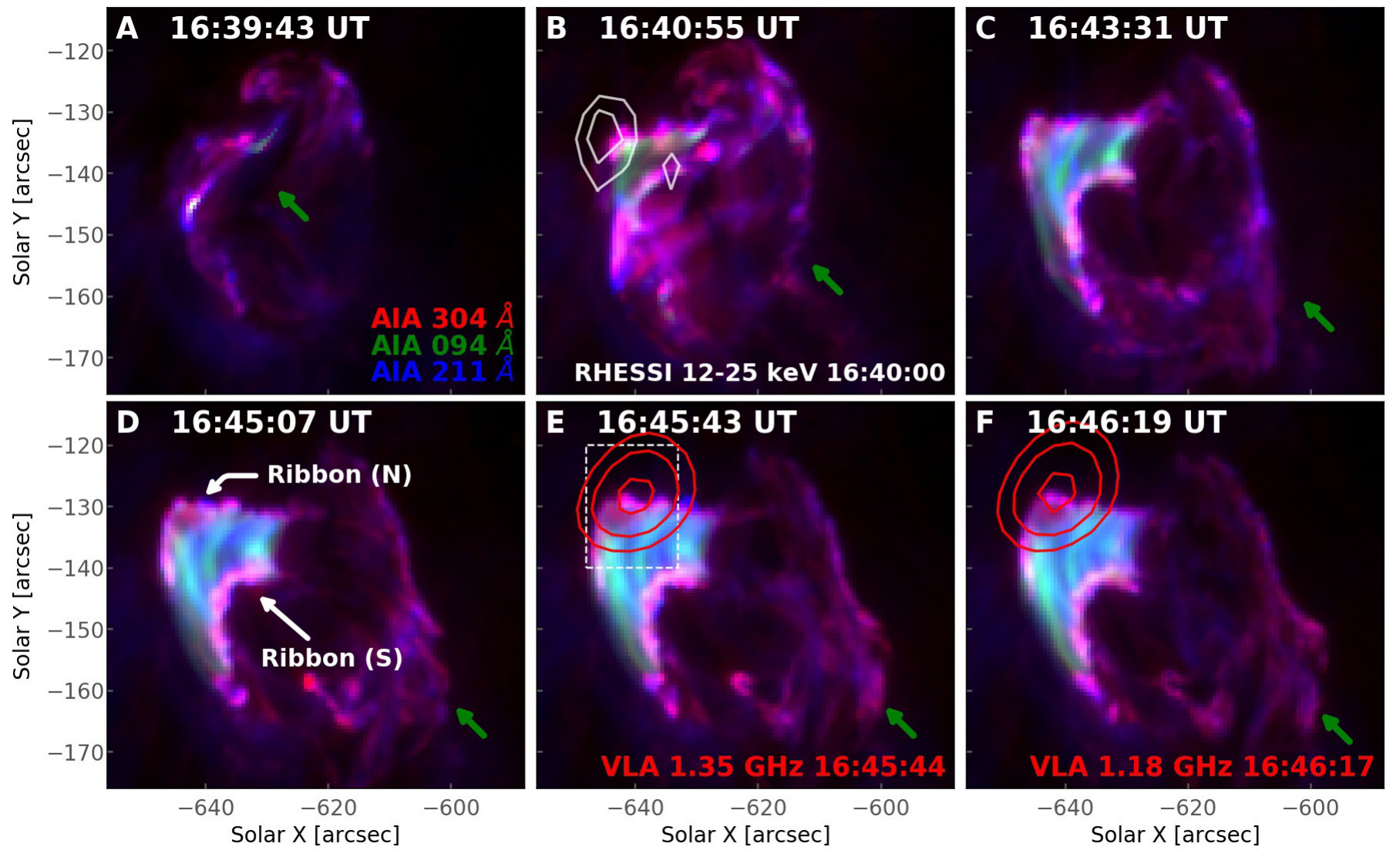}
\caption{Composite EUV image series of \textit{SDO}/AIA 304 \AA\ (red), 94 \AA\ (green), and 211 \AA\ (blue) EUV bands. The radio sources are shown in red contours at 91 \%, 95 \% and 99 \% of the maximum. White contours are 60 seconds integration of 12-25 $\mathrm{keV}$ HXR emission by \textit{RHESSI} during the flare's early impulsive phase. The erupting filament is indicated in (A)--(F) by green arrows, and the two flare ribbons are marked by white arrows in (D). An animation of this figure is available in the online
journal. The animation runs from 16:38 to 16:51 UT, and includes the \textit{RHESSI} and VLA annotations shown in (B), (E), and (F) of the static figure. (An animation of this figure is available.) \label{fig-euv}} 
\end{figure*}

Figures \ref{fig-overview} and \ref{fig-euv} show an overview of the time history and general context of the flare event. The \textit{GOES} 1--8 \AA\ SXR flux starts to rise at 16:39 UT and peaks at around 16:46 UT, during which time a filament is seen to erupt (green arrows in Figure \ref{fig-euv}) but it does not fully detach from the surface and forms a coronal mass ejection---a phenomenon known as a ``failed eruption.'' During this period, both the HXR light curve (blue curve in Figure \ref{fig-overview}(D)) and the SXR derivative (red curve in Figure \ref{fig-overview}(D)) display multiple bursty features, which is characteristic of the flare's impulsive phase, during which the primary energy release occurs. Precipitating nonthermal electrons lose most of their energy in the dense chromosphere, resulting in HXR sources at the footpoints of the reconnected flare arcades via bremsstrahlung radiation (contours in Figure \ref{fig-euv}(B)). Bright flare ribbons, visible in UV/EUV passbands (shown in Figure \ref{fig-euv} in purple, which is mostly contributed by AIA 304 \AA), are formed due to heating of the chromospheric/photospheric material by precipitated nonthermal electrons or by other means. The evaporated chromospheric material fills the flare arcades and forms bright coronal loops, best seen in EUV passbands that are sensitive to relatively high coronal temperatures (green and blue colors in Figure \ref{fig-euv}, which show AIA 211 and 94 \AA\ bands that correspond to plasma temperatures of 2 MK and 7 MK, respectively). Many of the impulsive peaks in the SXR derivative have counterparts in the light curves from the Radio Solar Telescope Network (RSTN) (Figure \ref{fig-overview}(B) and (C)), which are also visible in the VLA 1--2 GHz dynamic spectrum as short-duration radio bursts (Figure \ref{fig-overview}(A)), suggesting that they are both closely associated with accelerated nonthermal electrons. The \dml\ bursts have complex fine spectrotemporal structures, especially in the lower-frequency portion of the radio dynamic spectrum.

\begin{figure*}[htb!]
\includegraphics[width=\textwidth]{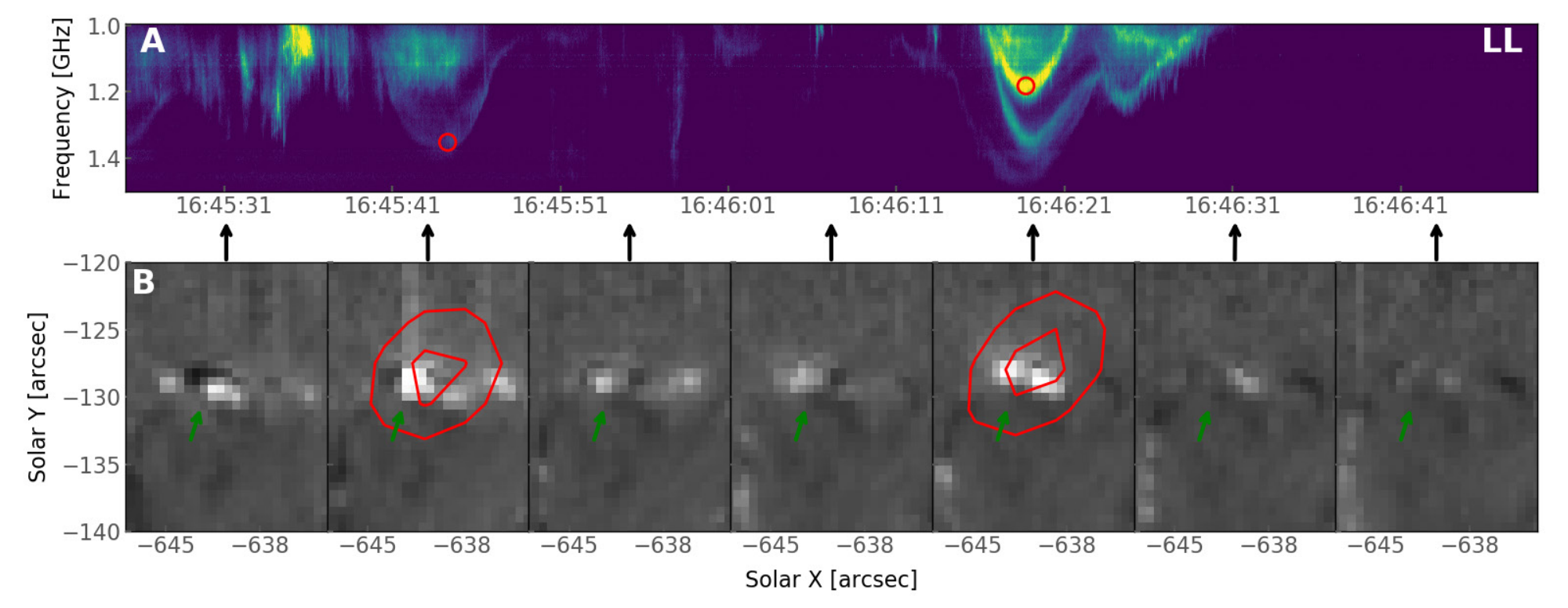}
\caption{\edit1{(A) VLA cross-power dynamic spectrum in LCP at 1--1.5 GHz obtained from a short baseline. (B) AIA 304 Å background-detrended image sequence at times marked by the black vertical arrows in (A), showing the EUV ribbon brightenings near the radio sources. Red contours are the radio images that correspond to Bursts 1 and 2 (the time and frequency are marked in the dynamic spectrum of (A) as red circles; contour levels are 97.5 \%, 99.5 \% of the image maximum). Green arrows indicate the location of the transient EUV brightening on the north ribbon. The field of view (FOV) is indicated by the dashed box in Figure \ref{fig-euv}(E)}. \label{fig-ribbonbrightenings}}
\end{figure*}

The radio bursts under study appear during the late impulsive phase (shaded area in Figure \ref{fig-overview}(A--C) demarcated with vertical dashed lines). Two main episodes can be distinguished in the dynamic spectrum, each of which lasts for $\sim$10--20 seconds (referred to as ``Burst 1'' and ``Burst 2'' hereafter). An enlarged view of these bursts is available in Figures \ref{fig-ribbonbrightenings}(A) and \ref{fig-radio-spec-imaging}(A). From the imaging data, the bursts have a peak brightness temperature $T_B$ of $\sim1.1 \times 10^7$ K. \edit1{The total flux density is $\sim1$ sfu (solar flux unit; 1 sfu $= 10^4$ Jy).} In addition, the bursts are nearly 100\% polarized with left-hand circular polarization (LCP). These properties are consistent with radio emission associated with a coherent radiation mechanism. In the dynamic spectrum, the bursts appear as arch-shaped emission lanes, which display a low-high-low frequency drift pattern. The frequency drift rate $d\nu/dt$ is between 60 and 200 MHz/s (or a relative drift rate of $\dot{\nu}/\nu\approx0.04$--0.2), which is about one order of magnitude lower than type III radio bursts emitted by beams of fast electrons, but similar to fiber bursts and lace bursts in the same frequency range \citep{1998A&A...333.1034B,2001A&A...375..638K,2013A&A...550A...1K,2007SoPh..245..327R,2017ApJ...848...77W}. Such bursts with an intermediate frequency drift rate are sometimes referred to as the ``intermediate drift bursts'' \citep{1998A&A...333.1034B,2005A&A...435.1137A,2006SoPh..237..153K}. The multiple episodes of positive- and negative-drifting features resemble to some extent the ``lace bursts'' in the literature \citep{2001A&A...375..638K,2005ApJ...631..612B,2012ApJ...745..186H}. However, the emission lanes of these bursts appear to be much smoother, while the lace bursts, at least from the few cases reported in the literature, have a much more fragmentary and chaotic appearance. 

Radio imaging of the bursts places the burst source (red contours in Figures \ref{fig-euv}(E) and (F)) near the northern flare ribbon. The location of the radio bursts is also very close to the \textit{RHESSI} 12--25 keV HXR footpoint source, shown in Figure \ref{fig-euv}(B) as white contours, albeit the latter is obtained several minutes earlier (at 16:40 UT) before the spacecraft enters the SAA. \edit1{A more detailed investigation reveals a close temporal and spatial association between the radio bursts and the transient (E)UV brightenings at the northern flare ribbon. Figure \ref{fig-ribbonbrightenings}(B) shows an AIA 304 \AA\ background-detrended image sequence during the time interval of the radio dynamic spectrum shown in Figure \ref{fig-ribbonbrightenings}(A). During this period, the northern ribbon features the appearance of two transient EUV brightenings during radio Bursts 1 and 2, and the location of the brightenings is very close to the radio source (red).} The appearance of the radio source during the flare's impulsive phase, as well as its close spatial and temporal association with the ribbon brightenings, suggests that the radio source is intimately related to the release and transport of the flare energy. More detailed discussions of the spectral, temporal, and spatial features of the bursts based on radio dynamic imaging spectroscopy will be presented in the next subsection.

\begin{figure*}[ht]
\includegraphics[width=\textwidth]{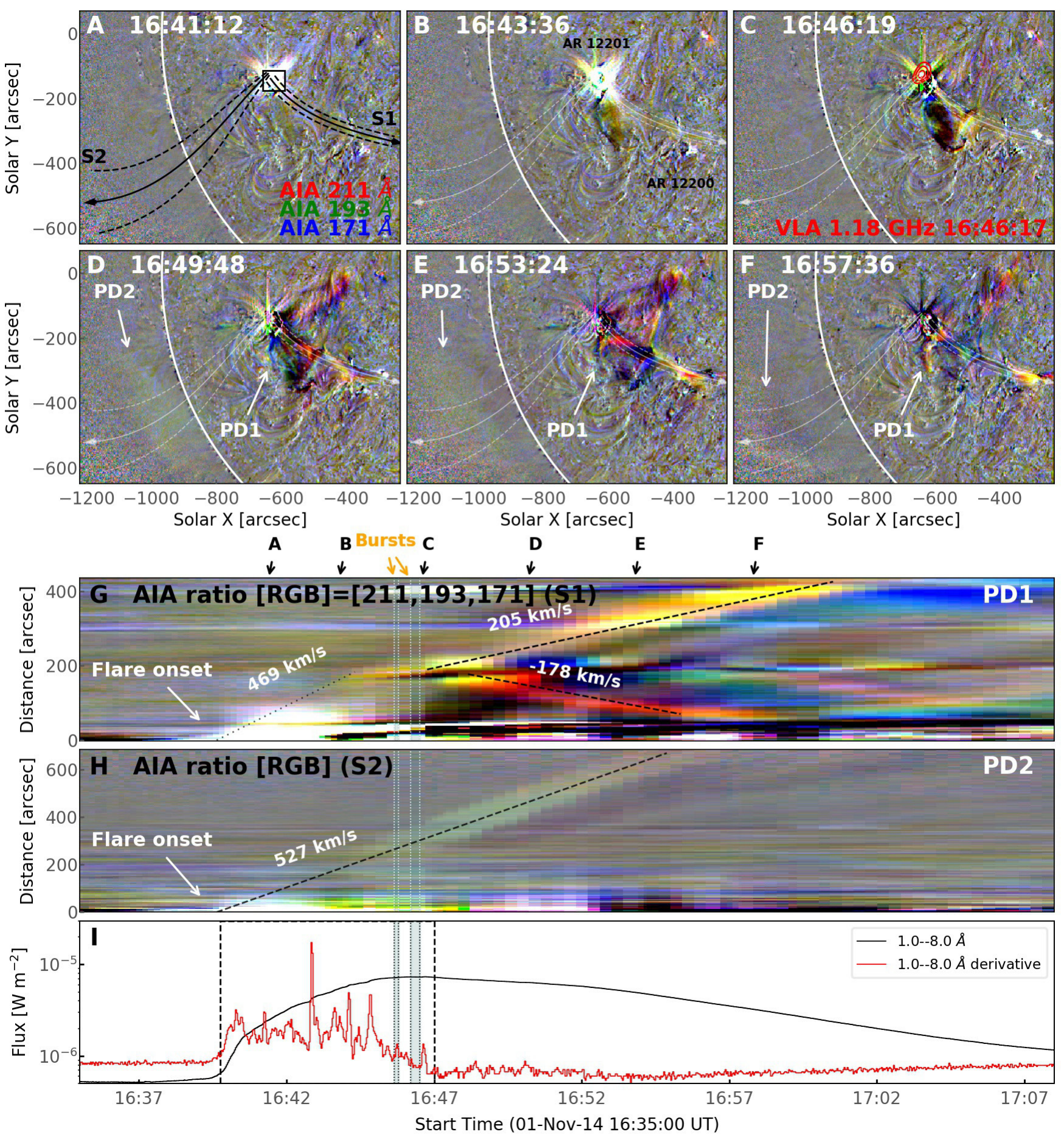}
\caption{Large-scale \edit1{propagating} disturbances observed by \textit{SDO}/AIA. (A--F) Snapshots of composite  AIA 211, 193 and 171 \AA\ running-ratio images. The corresponding times of the snapshots are marked by the black arrows above the time-distance plots in (G) and (H) obtained along two slices in (A) labeled ``S1'' and ``S2'' that follow the propagation direction of the two large-scale waves. The radio source in (C) is shown as red contours (50 \%, 70 \% and 90 \% of the maximum). The field of view of the EUV images in Figure \ref{fig-euv} is indicated by a black box in 
(A). \textit{GOES} 1--8 \AA\ SXR light curve of the flare and its time derivative are shown in (I). The periods of the two episodes of the radio bursts under study are demarcated by vertical lines in (G-I). The animation shows the \textit{SDO}/AIA running-ratio images from 16:35 to 17:07 UT on the left panel and the synchronized time-distance plots on the right. The
annotations are the same as in the static figure.
(An animation of this figure is available.) \label{fig-LSWave}}
\end{figure*}

Another interesting feature of this event is that it is accompanied by large-scale, \edit1{fast-propagating} disturbances \edit1{(``PDs'' hereafter), observed in EUV,} during  the impulsive and gradual phase of the flare; they are usually interpreted as propagating MHD waves in the corona \citep{2012SoPh..281..187P,2013ApJ...776...58N,2014SoPh..289.3233L,2015LRSP...12....3W,2017SoPh..292....7L,2018ApJ...864L..24L}. Using AIA 171, 193, and 211 \AA\ running-ratio images (ratio of current frame to the previous frame), a large-scale \edit1{PD feature} (denoted as \edit1{``PD1''} in Figure \ref{fig-LSWave}) is present in the area between AR 12201 and AR 12200. In addition, another large-scale \edit1{PD} appears to move outward above the limb (denoted \edit1{``PD2''} in Figure \ref{fig-LSWave}). The temporal evolution of the two \edit1{PDs} is displayed in the time-distance plots in Figure \ref{fig-LSWave}(G) and (H), made along two slices labeled ``S1'' and ``S2'' in Figure \ref{fig-LSWave}(A), respectively. The initialization of the large-scale \edit1{PDs} coincides with the onset of the flare, demonstrating their close association with the flare energy release. The large-scale \edit1{PDs} propagate at a speed of 400--500 km s$^{-1}$,  with \edit1{PD1} clearly experiencing multiple deflections by magnetic structures of the ARs. We note that the radio bursts are observed during the period when \edit1{PD1} remains in the flaring region (Figure \ref{fig-LSWave}(C)). This is a strong indication of the presence of ubiquitous MHD disturbances in and around the flaring region during the time of the radio bursts.

\subsection{Radio Dynamic Spectroscopic Imaging}\label{sec-spec-imaging}

\begin{figure*}[htb!]
\includegraphics[width=\textwidth]{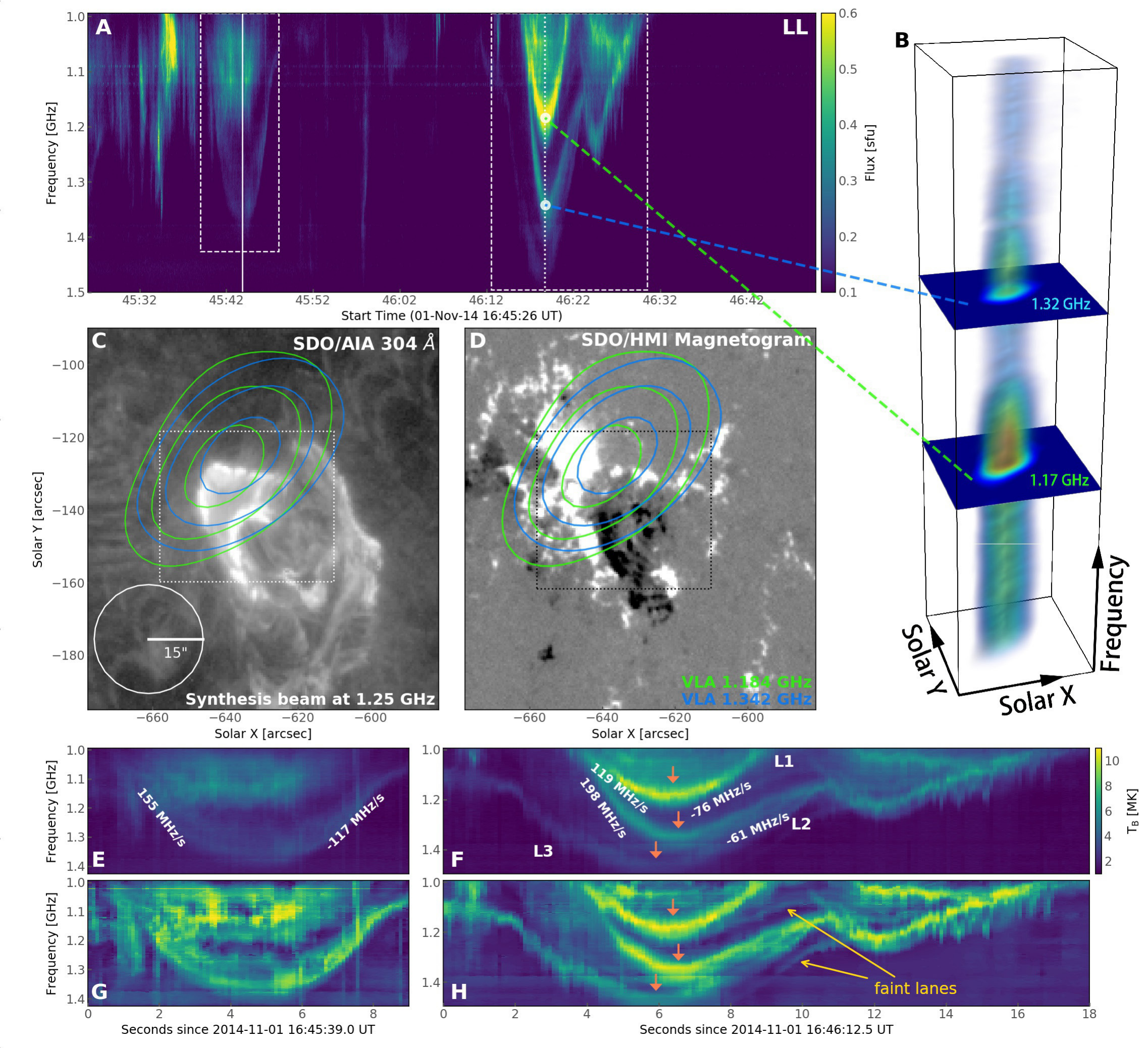}
\caption{(A) VLA cross-power dynamic spectrum in LCP at 1--1.5 GHz obtained from a short baseline. The vertical solid line denotes the frequency turnover time of Bursts 1 and 2. (B) Three-dimensional rendering of the spectral image cube of an 100 ms integration snapshot at 16:46:18.2 UT. 256 independent frequency channels are shown. The two horizontal slices indicate radio images at two selected frequencies that correspond to the two intensity peaks at the particular time, shown also in (C) as colored contours overlaid on the \textit{SDO}/AIA EUV 304 \AA\ image at 16:46:19 UT. The contour levels are at 50\%, 70\% and 90\% of the maximum. The circle in (D) represent the size of the restoring beam. (D) Same as (C), but the HMI line-of-sight magnetogram is shown as the background. (E) and (F) Spatially-resolved vector dynamic spectrum of the two burst episodes. (G) and (H) Feature-enhanced version of the dynamic spectra in (E) and (F). \label{fig-radio-spec-imaging}}
\end{figure*}

The capability of simultaneous imaging and dynamic spectroscopy offered by the VLA allows each pixel in the dynamic spectrum to form a radio image. As an example, Figure \ref{fig-radio-spec-imaging}(B) shows a three-dimensional (3D) rendering of a VLA spectral image cube taken for Burst 2 within a 100 ms integration (at 16:46:18.2 UT; the timing is shown as the vertical dotted line in panel (A)). The two horizontal slices in Figure \ref{fig-radio-spec-imaging} (B) indicate the radio images at the peak frequencies of the two emission lanes at that time (circles in panel (A)). The same two radio images are shown in Figures \ref{fig-radio-spec-imaging}(C) and (D) as green and blue contours overlaid on the AIA EUV 304 \AA\ image and the HMI photospheric line-of-sight (LOS) magnetogram respectively. As discussed in the previous subsection, the radio sources are located near the northern flare ribbon. In the magnetogram, this flare ribbon corresponds to a region with a positive magnetic polarity. As the bursts are 100\% LCP, they are likely polarized in the sense of $o$ mode.

We produce an independent 3D spectral image cube for each time pixel when the radio burst of interest is present in the radio dynamic spectrum, thereby creating a four-dimensional (4D) spectrotemporal image cube. From the 4D cube we are able to derive the spectrotemporal variation intrinsic to this radio source of interest by isolating its flux from all other sources present on the solar disk in the spatial domain, resulting in a spatial resolved, or ``vector'' radio dynamic spectrum of the source (Figure \ref{fig-radio-spec-imaging}(E) and (F)). This technique was first introduced by \citet{2015Sci...350.1238C} based on VLA dynamic spectroscopic imaging data, and was subsequently applied in a number of recent studies with VLA data \citep{2017ApJ...848...77W,2018ApJ...866...62C}. A similar approach is discussed in a recent study by \citet{2017SoPh..292..168M} based on data from the Murchison Widefield Array. The resulting vector dynamic spectra show clearer features of the radio bursts than the cross-power dynamic spectra obtained at short baselines (which are a proxy for the total-power dynamic spectra; Figure \ref{fig-radio-spec-imaging}(A)). The improvement, however, is not substantial, which is consistent with the imaging results in which this burst source is shown as the dominant emission on the solar disk. To highlight the fine structure of the bursts, we further enhance the vector dynamic spectrum by using the contrast-limited adaptive histogram equalization technique \citep{Pizer1987}, shown in Figures \ref{fig-radio-spec-imaging}(G) and (H). 

Bursts 1 and 2 share similar spectrotemporal features. They contain at least one emission lane that starts with a positive drift rate toward higher frequency ($d\nu/dt>0$, sometimes referred to in the literature as ``reverse drift'' because ``normal drift'' bursts show negative frequency drifts). It then turns over rather smoothly at the highest frequency point and drifts toward lower frequency with a negative frequency drift rate ($d\nu/dt<0$). The total frequency variation $\Delta \nu_{\rm tot}/\nu$ can be up to 30\%. Burst 2 undergoes two repeated cycles of positive-to-negative frequency drift. At least three distinct emission lanes are clearly visible (denoted as ``L1'', ``L2'', and ``L3'' in Figure \ref{fig-radio-spec-imaging}(F)) with two additional faint lanes that can only been distinguished in the enhanced dynamic spectrum (arrows in Figure \ref{fig-radio-spec-imaging}(H)). Although the three bright emission lanes of Burst 2 occur close together in time, they differ in their intensity, peak emission frequency, frequency drift rate, and frequency turnover time $t_{\rm o}$ (defined as the time when the emission frequency reaches the highest value and the frequency drift rate $\dot{\nu}$ goes to nearly zero; it is indicated by orange arrows in Figure \ref{fig-radio-spec-imaging}(F) and (H)). The average instantaneous frequency bandwidth $\Delta \nu$ of the emission lanes is about 60--100 MHz, corresponding to a relative frequency bandwidth $\Delta \nu/\nu\approx 6\%$.  

\begin{figure*}[ht]
\includegraphics[width=\textwidth]{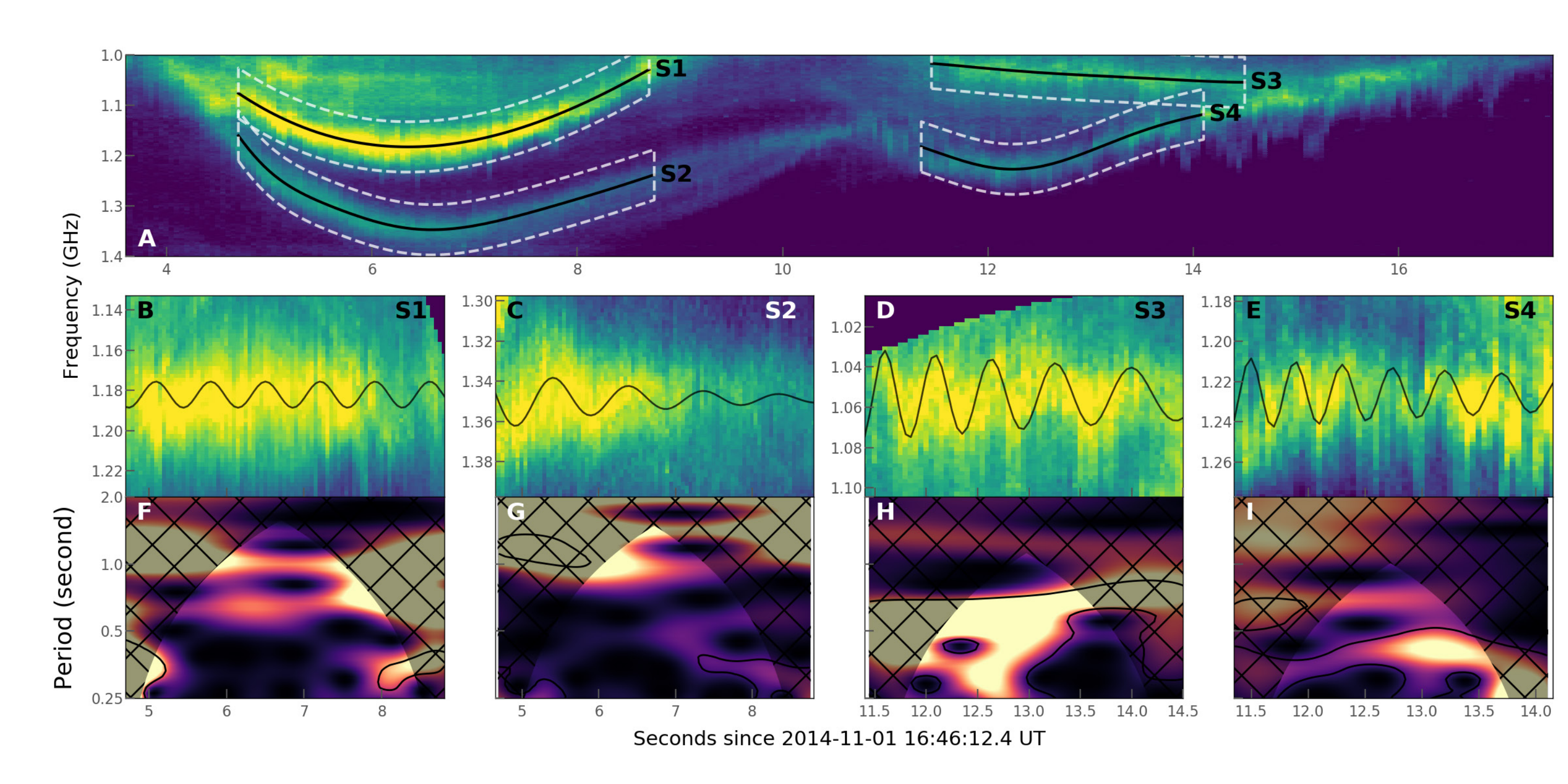}
\caption{Subsecond-scale oscillations in the emission lanes of the \dml\ bursts. (A) Dynamic spectrum  of Burst 2 with the full 50-ms time resolution. Four segments of emission lanes, labelled ``S1'', ``S2'', ``S3'', ``S4'', are selected for in-depth analysis shown in (B)--(I). (B)--(E) Detrended dynamic spectra of the four segments S1--S4. (F--I) Wavelet power spectra of the oscillations for the four segments S1--S4.} \label{fig-wavelet}
\end{figure*}

More detailed inspection of the dynamic spectral features of the stronger burst (Burst 2; based on the full 50 ms cadence data) reveals multitudes of very short, subsecond-scale fine structures on each emission lane (Figure \ref{fig-wavelet}(A)). Figures \ref{fig-wavelet}(B--E) provide an enlarged view of four segments of the emission lanes for Burst 2 (labelled as ``S1'', ``S2'', ``S3'', ``S4'' in Figure \ref{fig-wavelet}(A)), which have been detrended to remove their overall frequency drift pattern. The bursts appear to oscillate quasi-periodically in their emission frequency around the central ``ridge'' of the emission lane. \edit1{We use a damped oscillation profile
\begin{equation}
\delta\nu(t) = \delta\nu_0\exp(-t/\tau_A)\sin\bigg[\frac{2\pi t}{P/(1-t/\tau_P)^3}\bigg]
\end{equation}
to fit the four segments (Figure \ref{fig-wavelet}(B--E)). The oscillations have an amplitude of $\delta\nu_0\approx10$--$30\ \rm{MHz}$ (or a relative amplitude of $\delta\nu/\nu$ of $\sim$1--2$\%$), period of $P\approx0.3$--1.0 s, and damping times of $\tau_A\approx0.5$--5 s in amplitude and $\tau_P \gtrsim$ 30 s in period.} Wavelet analysis of such oscillation patterns in emission frequency confirms that the oscillations display very short, subsecond-scale periods ranging from $\sim$0.3--1.0 s (Figure \ref{fig-wavelet}(F--I)).

Radio imaging of each pixel in the dynamic spectrum where the bursts are found provides key information on the spatial variation of the radio source as a function of time and frequency. For each image at a given frequency $\nu$ and time $t$, we fit the source with a 2D Gaussian function and determine the source centroid $I_{\rm pk}(\theta, \phi, \nu, t$), where $I_{\rm pk}$ is the peak intensity, and $\theta$ and $\phi$ are the centroid position in helioprojective longitude and latitude. As shown in several previous studies, the uncertainty of the centroid location for unresolved, point-like sources (which is likely the case for the coherent bursts under study) is determined by $\sigma\approx\theta_{\rm FWHM}/\mathrm{S/N}\sqrt{8\ln 2}$, where $\theta_{\rm FWHM}$ is the FWHM beamwidth and S/N is the ratio of the peak flux to the root-mean-square noise of the image \citep{1988ApJ...330..809R,1997PASP..109..166C,2018ApJ...866...62C}. In our data, typical values are $\theta_{\rm FWHM}\approx 30''$ and $\mathrm{S/N}\gtrsim20$, which give $\sigma\lesssim0.6''$. However, as discussed later in Section \ref{sec-discussion}, the bursts are likely associated with fundamental plasma radiation, which is known to be prone to scattering effects as the radiation propagates through the inhomogeneous corona toward the observer \citep{1994ApJ...426..774B,2017NatCo...8.1515K, 2018ApJ...856...73C,2018SoPh..293..132M}. Therefore, the estimate of uncertainty given above should only be considered as a lower limit. In fact, by obtaining the centroid locations of all frequency-time pixels on the emission lane within a small time period ($\sim$0.5 s) and frequency range ($\sim$50 MHz), we find that they are distributed rather randomly within an area of a FWHM size of $\sim 2''\times2''$. Hence we estimate the actual position uncertainty of the centroids as $\sigma\approx1''$. 

\begin{figure*}[htb!]
\includegraphics[width=\textwidth]{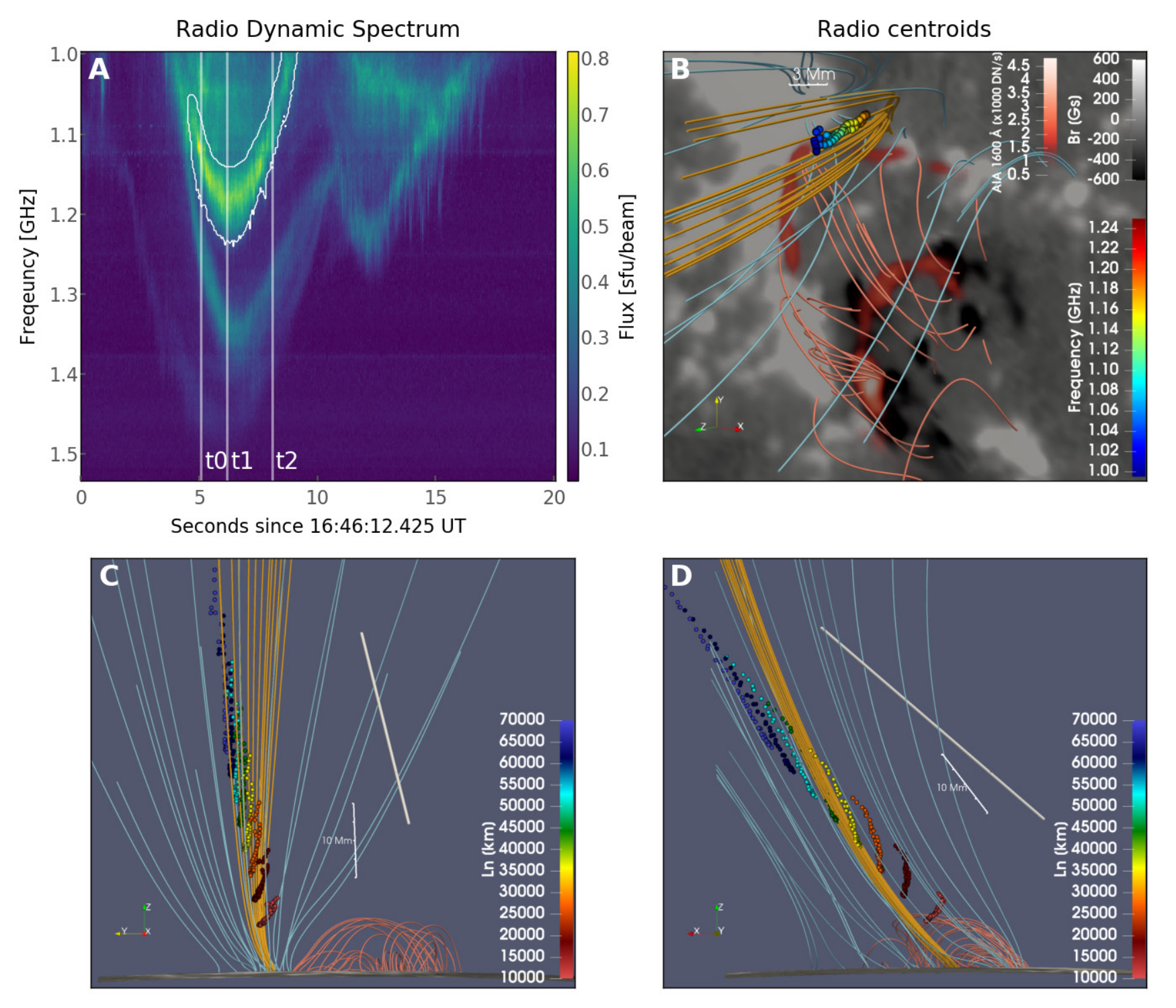}
\caption{Three-dimensional locations of the radio burst centroids for emission lane L1 in Burst 2 (as denoted in Figure \ref{fig-radio-spec-imaging}(F)). (A) Dynamic spectrum of the Burst 2. All the frequency--time pixels on emission lane L1 selected for dynamic spectroscopic imaging are enclosed by solid white curves. (B) Locations of burst centroids of the selected frequency--time pixels of L1. Each centroid represents the average location of all frequency pixels across the emission lane at a given time, colored according to their mean frequency  (frequency increases from blue to red). Colored curves are selected magnetic field lines from the potential field extrapolation model based on the \textit{SDO}/HMI line-of-sight photospheric magnetogram observed at around 17:00 UT (grayscale background). Pink field lines denote closed field associated with the reconnected loops, while yellow (cyan) ones denote open field lines that connect to the northern ribbon (sunspot). The double flare ribbons seen in AIA 1600 \AA\ are shown in red. The FOV is indicated by the dashed box in Figure \ref{fig-radio-spec-imaging}(C) and (D). (C) Three-dimensional distribution of the radio centroids assuming different density scale heights $L_n$, viewed from the east side of the active region. Each set of centroids in the same color represents the 3D projection of all the centroids shown in (B) by assuming a given value of $L_n$. The thick white line indicates the LOS of an Earth-based observer. (D) Same as (C), but viewed from the north side of the active region. \label{fig-L1-3d}}
\end{figure*}

We focus on Burst 2 for detailed investigations of the spatial, temporal, and spectral variation of the source centroid since it has the best S/N. For each emission lane, we first  extract all time and frequency pixels where the intensity exceeds 50\% of its peak intensity. An example for such a selection for lane L1 of Burst 2 is shown in Figure \ref{fig-L1-3d}(A) enclosed by the white contour. Figure \ref{fig-L1-3d}(B) shows the resulting centroid positions as a function of frequency (colored dots from blue to red in increasing frequency) for emission lane L1. To further improve positional accuracy and reduce cluttering in the figure, each dot in the plot represents the average position for centroids at all frequency pixels across the emission lane (that have an intensity above 50\% of the peak) for a given time $t$, with the color representing their mean frequency. The background of Figure \ref{fig-L1-3d}(B) is the HMI photospheric magnetogram shown in grayscale, overlaid with the AIA 1600 \AA\ image. The latter clearly shows the double flare ribbons in red. 

Figure \ref{fig-L1_time}(A) shows the same distribution of radio centroids derived from emission lane L1 as in Figure \ref{fig-L1-3d}(B), but colored in time instead. It displays an evident motion in projection: the radio source first moves toward the flare ribbon as frequency increases (blue to red color in Figure \ref{fig-L1-3d}(B) and blue to white color in Figure \ref{fig-L1_time}(A)) until it reaches the maximum frequency at the lowest height, and then bounces back in the opposite direction away from the ribbon as frequency decreases (red to blue color in Figure \ref{fig-L1-3d}(B) and white to red color in Figure \ref{fig-L1_time}(A)). The average speed in projection is $\sim$1000--2000 km s$^{-1}$, which is typical for propagating Alfv\'en or fast-mode magnetosonic waves in the low corona \citep[e.g.,][]{2013ApJ...776...58N,2018ApJ...864L..24L}.
This is a strong indication that the radio emission is associated with a propagating Alfv\'en or fast-mode MHD disturbance in a magnetic tube in the close vicinity of the flare ribbon. As discussed in Section \ref{sec-overview}, the presence of ubiquitous MHD disturbances in the flaring region is strongly implicated by the observation of large-scale, fast \edit1{PDs} observed by \textit{SDO}/AIA at about the same time. 

\subsection{Motion of Radio source motion in 3D}\label{sec-3d}

In order to place the location of the radio centroids into the physical context of the flare, we perform potential field extrapolation based on the \textit{SDO}/HMI LOS photospheric data right after the flare peak at 17:00 UT \citep{2014SoPh..289.3549B, 2014SoPh..289.3483H} to derive the coronal magnetic field. Selected magnetic field lines from the extrapolation results are shown in Figure \ref{fig-L1-3d}(B) for regions around the location of the radio burst centroids and the postflare arcades. It is shown that the spatial distribution of the positions of the radio centroids at different frequencies tends to follow the magnetic field lines (yellow) rooted around the northern flare ribbon, with its higher-frequency end located closer to the ribbon. This is consistent with the expectation for plasma radiation, in which emission occurs at a higher emission frequency in regions with higher plasma density, which are typically located at lower coronal heights ($\nu\approx s\nu_{pe}\approx 8980s\sqrt{n_e}$ Hz, where $s$ is the harmonic number, $\nu_{\rm pe}$ is the electron plasma frequency, and $n_e$ is the local electron density in cm$^{-3}$).

\begin{figure*}[htb!]
\includegraphics[width=\textwidth]{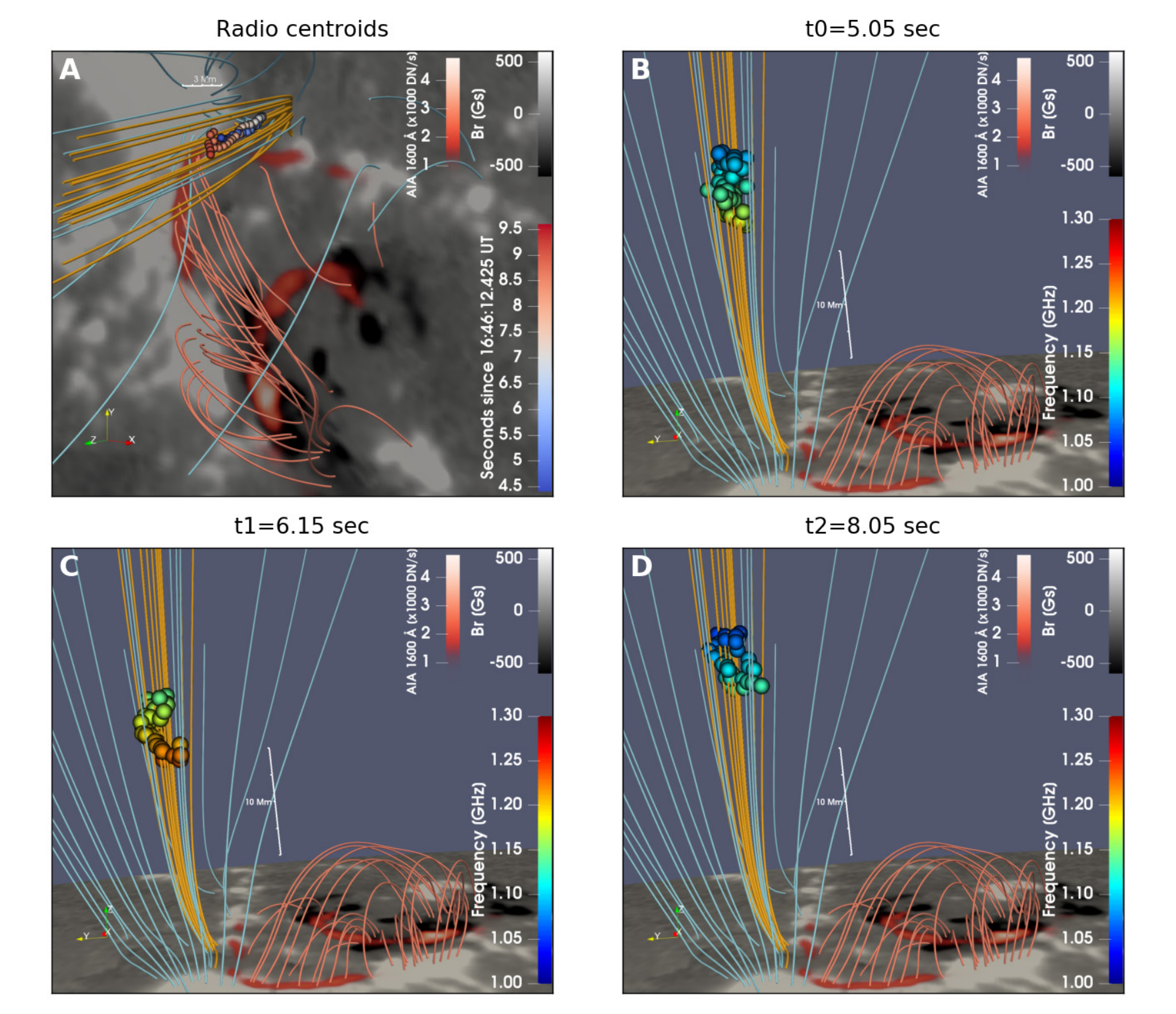}
\caption{(A) Similar as Figure \ref{fig-L1-3d}, but the centroids are colored from blue to red in time. (B-D) Three-dimensional views of emission centroids of L1 at three 100 ms integrations (denoted as t0, t1, and t2 in Figure \ref{fig-L1-3d}(A)). The centroids and contours are colored from blue to red in increasing frequency. An animation of this figure is available in the online journal. The animation includes the dynamic spectrum of Burst 2 from Figure \ref{fig-L1-3d} (left panel) synchronized to the motion of the three-dimensional distribution of the radio centroids of emission lane L1 against the potential field extrapolation model (right panel). The animation proceeds in 0.1 s increments from t = 4.55 to 9.15 s from 16:46:12.425 UT. (An animation of this figure is available in the online journal.) \label{fig-L1_time}}
\end{figure*}

Since the emission is highly polarized, it is reasonable to assume fundamental plasma radiation as the emission mechanism responsible (i.e., harmonic number $s$=1 and $\nu \approx \nu_{\rm pe}$). In this case, the emission frequency $\nu$ of the radio source centroid $I_{\rm pk}(\theta, \phi, \nu, t$) can be directly translated into the plasma density of the source $n_e$. By further assuming a coronal density model $n_e(h)$ where $h$ is the coronal height, we can thus map the measured centroid locations in 2D projection to three dimensional (3D) locations in the corona, i.e., from $I_{\rm pk}(\theta, \phi, \nu, t$) to $I_{\rm pk}(\theta, \phi, h, t$). A similar practice has been used in \citet{2005A&A...435.1137A}, and more recently \citet{2017ApJ...848...77W}, for deriving 3D trajectories of \dml\ fiber bursts in the corona. Here we adopt a barometric density model with an exponential form
\begin{equation}
 n_\mathrm{e}(h)=n_{e0}\,\mathrm{exp}\left(-\frac{h-h_0}{L_n}\right)   
\label{equ-density-model}
\end{equation}
where $h$ is the height above the solar surface, $L_n$ is the density scale height, and $n_{e0}$ is the density at a reference height $h_0$. Such a density model describes the variation in density for an isothermal, plane-parallel atmosphere under hydrostatic equilibrium (e.g., \citealt{2005psci.book.....A}), and has been widely used in the literature as a zero-order approximation for estimating the coronal heights of various solar coherent radio bursts \citep[e.g.,][]{2005A&A...435.1137A,2011ApJ...736...64C,2013ApJ...763L..21C,2017ApJ...848...77W}. For simplicity, we fix the parameters $n_{e0}$ and $h_0$ to be $\sim 3\times 10^{10}\ \mathrm{cm^{-3}}$ and $\sim$2000 km at the top of the chromosphere according to the VAL model \citep{1981ApJS...45..635V}, and investigate the effect of different choices of $L_n$ on the resulting 3D distribution of the radio source centroids.

\begin{figure*}[p]
\includegraphics[width=\textwidth]{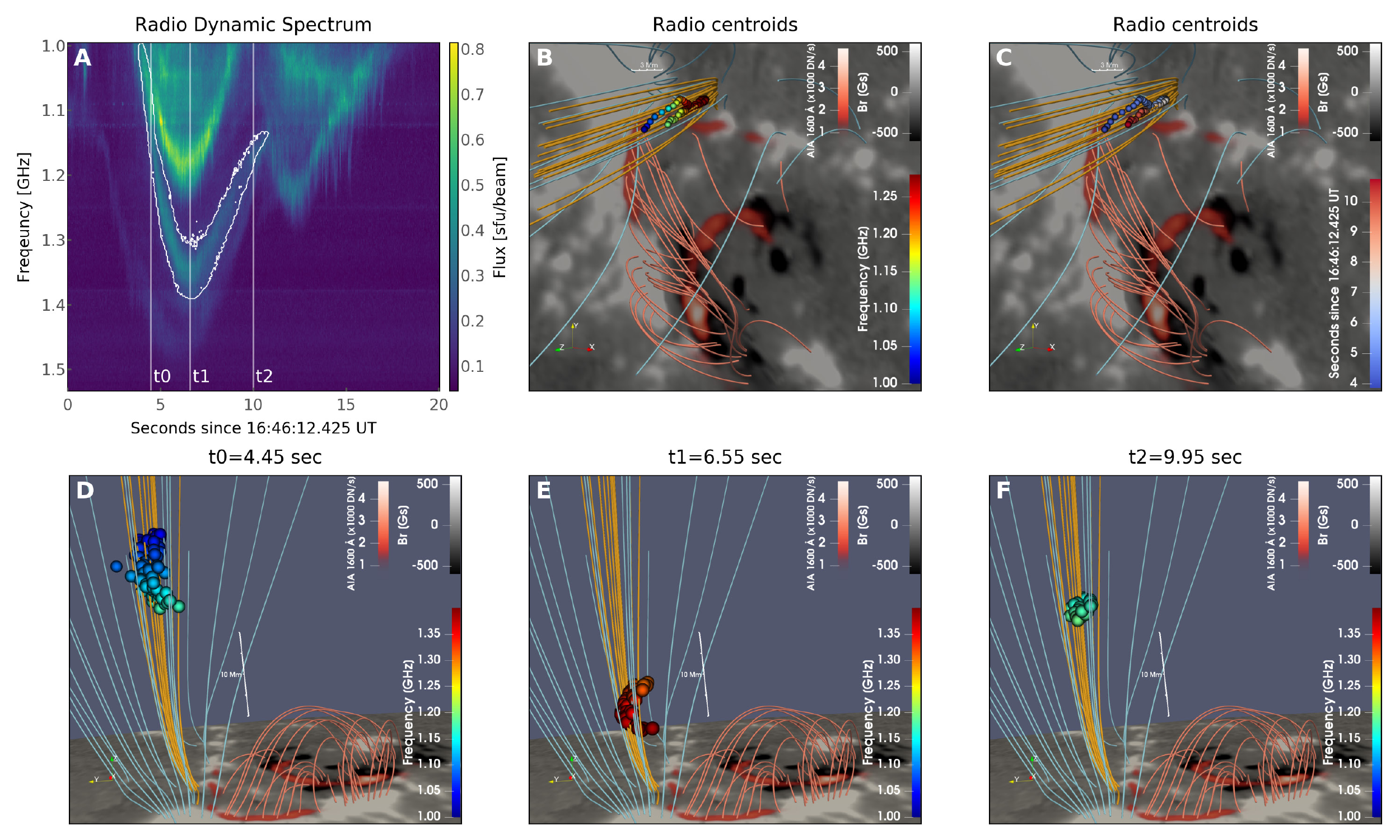}
\caption{(A) and (B) Same as Figure \ref{fig-L1-3d}(A) and (B) but for the emission lane L2 of Burst 2. (C-F) Same as Figure \ref{fig-L1_time}(A-D) but for the emission lane L2 of Burst 2. An animation of this figure is available in the online journal. The animation includes the dynamic spectrum of Burst 2 from (A) in the static figure (left panel)
synchronized to the motion of the 3D distribution of the radio centroids of emission lane L2 against the model of potential field extrapolation (right panel). The
animation proceeds in 0.1 s increments from t=3.85 to 10.75 s from 16:46:12.425 UT.
(An animation of this figure is available.) \label{fig-L2}}
\end{figure*}

\begin{figure*}[p]
\includegraphics[width=\textwidth]{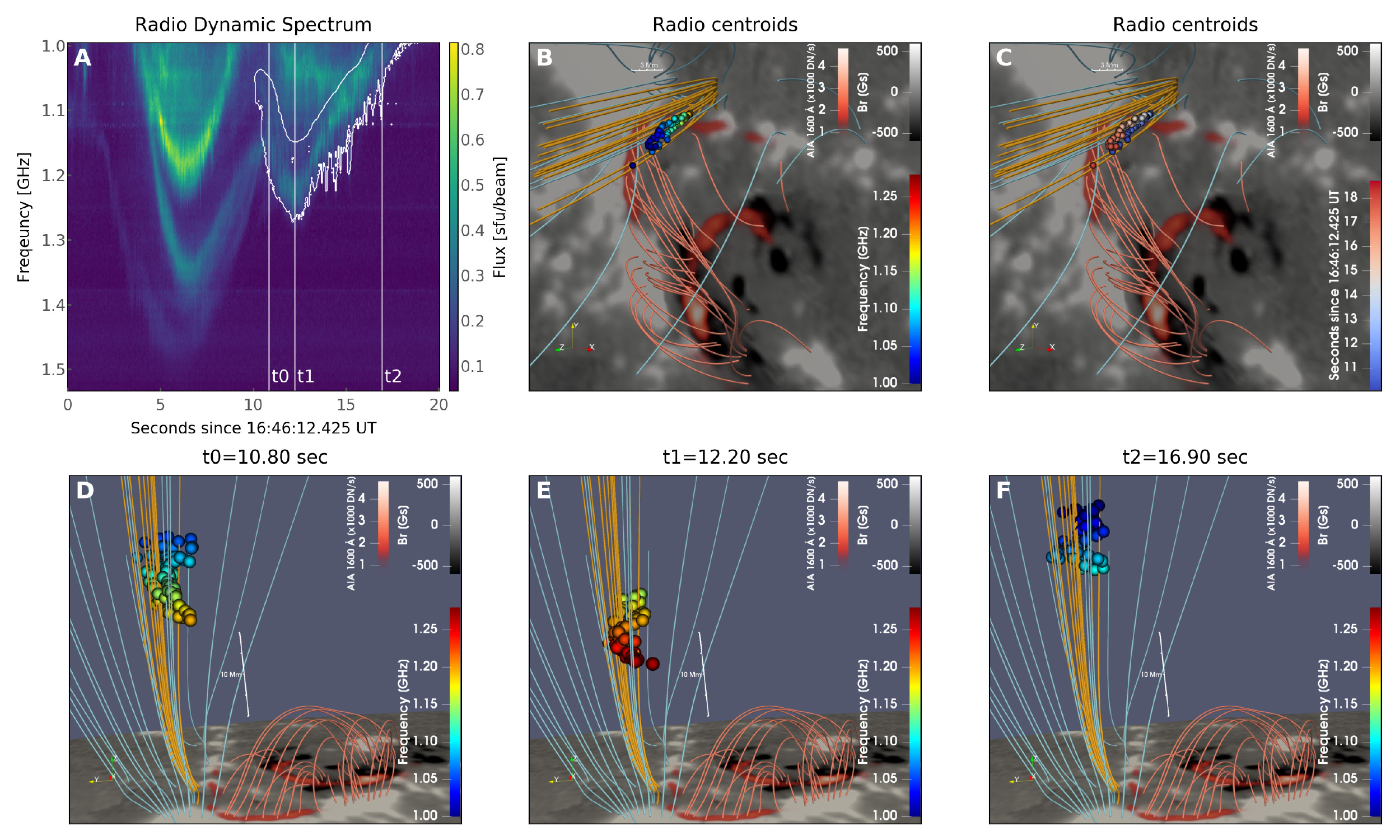}
\caption{Same as Figure \ref{fig-L2}, but for the emission lane L3 of Burst 2. An animation of this figure is available in the online journal. The animation includes the dynamic
spectrum of Burst 2 from (A) in the static figure (left panel) synchronized to the motion of the 3D distribution of the radio centroids of emission lane L3 against the
model of potential field extrapolation (right panel). The animation proceeds in 0.1 s increments from t=10.1 to 18.4 s from 16:46:12.425 UT.
(An animation of this figure is available.) \label{fig-L3}}
\end{figure*}

Figures \ref{fig-L1-3d}(C) and (D) demonstrate the inferred 3D distributions of the radio source centroids with different choices of $L_n$ from 10 Mm to 70 Mm, viewed from the eastern and northern sides of the AR respectively. Each set of 3D centroid positions at a given $L_n$ is shown as dots of the same color (from red to blue in increasing $L_n$). It is obvious from the figure that the choice of a greater value of $L_n$ yields a more stretched distribution of the radio centroids in height, and vice versa. Such a proportionality between the vertical extent $h_{\rm tot}$ of the radio sources and $L_n$ is straightforward to find if we combine the barometric density model (Eq. \ref{equ-density-model}) with the frequency--density relation for plasma radiation $\nu \propto \sqrt{n_e}$, which gives $h_{\rm tot} \approx 2L_n\Delta\nu_{\rm tot}/\nu$, where $\Delta\nu_{\rm tot}$ is the total frequency width of the radio burst determined from the dynamic spectrum. More importantly, different choices of $L_n$ affect how the radio source centroids are distributed with regard to the extrapolated magnetic field lines in 3D. For small $L_n$ values, the centroids tend to be distributed across the field lines within a small range of vertical heights, while for $L_n$ values in the intermediate range ($\sim$35--50 Mm), the spatial extent of the centroids tends to agree with the direction of the extrapolated field lines. As discussed earlier, the temporal evolution of a radio source (1000--2000 km s$^{-1}$ in projection) is consistent with a physical motion of the emission source at Alfv\'enic or fast-mode magnetosonic speed. Because the radio source appears to be closely associated with the flare ribbon both spatially and temporally (see Section \ref{sec-overview}), we assume that the radio source moves along (or within a small angle with respect to) the reconnected magnetic loops that link to the flare ribbon. In this case, the corresponding $L_n$ values fall into the $\sim$35--50 Mm range. For subsequent discussions, we will adopt $L_n=40$ Mm, with the understanding that this parameter is not very well determined due to the inherent limitations of magnetic field extrapolation the uncertainty on the exact direction of propagation of the radio source in 3D, and it may vary from burst to burst. 

Figure \ref{fig-L1_time}(B)--(D) shows the inferred 3D spatial and temporal evolution of the radio centroids of emission lane L1 after adopting the coronal density model with $L_n=40$ Mm, viewing from the east side of the AR. It is clear that the radio source first moves downward along a converging magnetic field tube (panels B and C) and then bounces backward after it reaches the lowest altitude (or highest frequency). We also extend the same analysis to emission lanes L2 and L3 of Burst 2. The results show a similar spatiotemporal evolution of the radio source centroids as lane L1 (Figures \ref{fig-L2} and \ref{fig-L3}). \edit1{We caution that the absolute height of the radio source as well as the point of reflection, however, depends strongly on the selection of parameters in the coronal density and magnetic field model adopted here}, which may very well be different for radio bursts propagating along different flare loops. Therefore, the 3D source evolution shown in Figures \ref{fig-L1_time}--\ref{fig-L3} should only be considered as a qualitative representation.

\section{DISCUSSIONS}\label{sec-discussion}
We briefly summarize the observational results in the previous section as follows. 
\begin{enumerate}
    \item The radio bursts of interest appeared during the late impulsive phase of a C7.2 two-ribbon solar flare that was associated with a failed filament eruption, when large-scale, fast-propagating EUV disturbances were observed throughout the flare region.
    \item The location of the radio source coincides with the northern flare ribbon and HXR footpoints. \edit1{In addition, the radio source appears to show close spatial and temporal association with transient (E)UV brightenings on the ribbon.}
    \item The bursts have a high brightness temperature of $>10^7$ K and are completely polarized in the sense of $o$ mode.
    \item The bursts consist of multiple emission lanes that exhibit a low--high--low frequency drift pattern in the radio dynamic spectrum with a moderate relative frequency drift rate of $\dot{\nu}/\nu \lesssim 0.2 \rm{s}^{-1}$, which is typical for intermediate drift bursts in the decimetric wavelength range.
    \item Imaging at all time and frequency pixels where the bursts are present shows that the radio source propagates at a speed of 1--2 Mm s$^{-1}$ in projection. The low--high--low frequency drift pattern corresponds to the source firstly moving downward along the flaring loop before it reaching the lowest point and bouncing back upward.
    \item Some of the emission lanes consist of multitudes of subsecond-period oscillations in emission frequency with an amplitude of \edit1{$\delta\nu/\nu\approx$1--2$\%$}.
\end{enumerate}

What is the nature of the propagating radio source that is reflected at or near the flare ribbon? First, it is most likely associated with fundamental plasma radiation, which is due to the nonlinear conversion from plasma Langmuir waves induced by the presence of nonthermal electrons. This is because that the bursts have a narrow frequency bandwidth ($\delta \nu/\nu \approx 6\%$) and fast temporally varying features, and are nearly 100\% polarized. Second, the propagation speed of the emission source (1--2 Mm s$^{-1}$ in projection) is too slow for electron beams emitting type III bursts (which usually propagate at 0.1$c$--0.5$c$, see, e.g., \citealt{2013ApJ...763L..21C, 2018ApJ...866...62C, 2017ApJ...851..151M}), but \edit1{likely} too fast for slow-mode magnetosonic waves, \edit1{unless the temperature in the source reaches over $\sim$50 MK}. 
The most probable candidate for the radio-emission-carrying disturbance is thus Alfv\'enic or fast-mode magnetosonic waves, which propagate at $\sim$1--4 Mm s$^{-1}$ under typical coronal conditions. The Alfv\'enic or fast-mode waves can be excited by a broadband driver, such as the impulsive flare energy release, and propagate outward from the site of energy release. For fast-mode waves to achieve focused, field-aligned energy transport, an overdense magnetic tube would be required to act as a waveguide \citep{1983Natur.305..688R,1984ApJ...279..857R,2004MNRAS.349..705N,2013A&A...558A..76R,2018ApJ...861...33K}, which, in our case, can be the freshly reconnected flaring loops that connect to the flare ribbons. The observed reflection of the waves at or near the flare ribbon may be due to sharp gradients at and/or below the transition region \citep{1982SoPh...80...99E,2008ApJ...675.1645F,2013A&A...558A..76R,2013ApJ...765...81R,2018ApJ...853..101R}. However, this is less clear from our observations regarding the physical connection between the nonthermal electrons (responsible for the production of Langmuir waves) and the MHD waves: the energetic electrons could be accelerated locally within the waves by a variety of means, including acceleration by parallel electric field, turbulence, or a first-order Fermi process with the wavefront acting as a moving mirror (e.g., \citealt{2008ApJ...675.1645F}), or they could originate from an acceleration site elsewhere (e.g., at the reconnection site or the top of the flare loops) but be trapped with the propagating MHD waves. 

It is particularly intriguing that some of the emission lanes show fast, subsecond-scale quasi-periodic oscillations in the emission frequency with an amplitude of $\delta\nu/\nu\approx$\edit1{1--2$\%$}. Under the plasma radiation scenario, $\delta\nu/\nu$ can be directly translated into small density perturbations of $\delta n_e/n_e\approx 2\delta\nu/\nu\approx $\edit1{2--4$\%$}. \edit1{If these small-amplitude oscillations in frequency can be interpreted as weak density perturbations associated with the propagating waves, the scenario of fast-mode magnetosonic mode scenario would be more probable, because pure Alf\'ven modes are incompressible.} We note that such small density disturbances are hardly detectable by current EUV or SXR imaging instrumentation, mainly because the resulting small fluctuation level in the EUV/SXR intensity $\delta I/I\lesssim 2\delta n_e/n_e\approx 4\%$ \citep{2003A&A...409..325C} is very difficult to detect against the background. In addition, the subsecond periodicity of the density perturbations is at least an order of magnitude below the time cadence of the current EUV/SXR imaging instrumentation (e.g., 12 s for \textit{SDO}/AIA). We note, however, that subsecond-scale oscillations in the solar corona have been reported in the literature based on non-imaging radio or X-ray light curves or dynamic spectra during flares \citep[e.g.,][]{1970A&A.....9..159R,1983SvAL....9..163B,1990SoPh..130..161F,1996SoPh..163..383Q,2007SoPh..246..431C,2010ApJ...723...25T,2013ApJ...777..159Y}. \citet{1987SoPh..111..113A} summarized the possible mechanisms into three categories: (1) quasi-periodic injections of nonthermal electrons, (2) fast cyclic self-organizing systems of plasma instabilities associated with the wave--particle or wave--wave interaction processes, and (3) MHD oscillations. While we cannot completely rule out the other possibilities, the observed oscillations in radio emission frequency (or plasma density), combined with the fast-moving radio source with a speed characteristic of Alfv\'enic or fast-mode waves, are a strong indication of a weakly compressible, propagating MHD wave packets in the flaring loops that cause localized quasi-periodic modulations of the plasma density along their way. 

The spatial scale of the radio-emitting fast-wave packages can be inferred from the instantaneous frequency bandwidth $\Delta \nu/\nu$ of individual emission lanes based on the plasma radiation scenario: $\Delta L = 2L_n(\Delta \nu/\nu)$, where $L_n=n_e/(dn_e/dl)$ is the density scale height. For a magnetic loop under hydrostatic equilibrium, the density gradient is along the vertical direction $z$, and the density scale height is $L_n=n_e/(dn_e/dl)=2k_BT/(\mu m_Hg)\approx46 T_{\rm MK}$ Mm, where $g$ is the gravitational acceleration near the solar surface, $m_H$ is the mass of the hydrogen atom, $T_{\rm MK}$ is the coronal temperature in megakelvin, and $\mu\approx 1.27$ is the mean molecular weight for typical coronal conditions \citep{2005psci.book.....A}. In this case, a frequency bandwidth of $\Delta \nu/\nu \approx 6\%$ implies a vertical extent of the source of $\Delta L_z \approx 5.5T_{\rm MK}$ Mm. Such an estimate of the source size is not inconsistent with the distribution on a small scale of a few megameters of the radio source centroids across all frequencies on the emission lane at a given time in the plane of the sky $\Delta L_{\parallel}$, although the latter is complicated by the scattering of the radio waves due to coronal inhomogeneities (see discussions in Section \ref{sec-spec-imaging}). It is interesting to note that this size estimation is about an order of magnitude smaller than the apparent size of each radio image (with a half-power-full-maximum size of $\sim$30--50 Mm; see Figure \ref{fig-radio-spec-imaging}(D)). Such an extended radio image can likely be attributed to the angular broadening of the radio source caused by random scattering of the radio waves traversing the inhomogeneous corona \citep{1994ApJ...426..774B}. Indeed, \citet{1994ApJ...426..774B} estimated an angular broadening of a few tens of arcseconds at our observing frequency and source longitude, which is of the same order of magnitude as our apparent source size. 

The wavelength associated with the subsecond-period oscillations can be estimated via $\lambda\approx v_{\rm p}P$, where $v_{\rm p}$ is the phase speed of the waves, taken to be of the same order of magnitude as the observed wave speed $\sim$3 Mm s$^{-1}$ (after assuming an inclination angle of 60$^\circ$ inferred from the extrapolation of the magnetic field, see Section \ref{sec-3d}) that presumably represents the group speed of the wave packet $v_{\rm g}$ (see, e.g., \citealt{1984ApJ...279..857R} for discussions regarding the relation between $v_{\rm p}$ and $v_{\rm g}$), and $P$ is the wave period, taken to be $\sim$0.5 s from the observed periods of the density fluctuations (see Figure \ref{fig-wavelet}). Therefore, the wavelength of the oscillations is estimated as $\lambda\approx 1.5$ Mm, much smaller than the size of the propagating radio source ($\Delta L > \Delta L_z \approx 5.5T_{\rm MK} $ Mm). We therefore argue that each radio source is likely a propagating MHD wave packet that consists of multiple short-period oscillations.

During each burst period, multiple emission lanes are present in the radio dynamic spectrum with almost synchronous frequency drift behavior (which is particularly clear for Burst 2; see Figure \ref{fig-radio-spec-imaging}(H)). Imaging results of the different emission lanes suggest that they are all located at the same site and share very similar spatiotemporal behavior in projection, but show subtle differences (see Figures \ref{fig-L1_time}(A), \ref{fig-L2}(C), and \ref{fig-L3}(C)). Their different emission frequencies, however, imply that the corresponding propagating disturbances have different plasma densities. Some other types of solar \dml\ bursts, in particular, zebra-pattern bursts (ZBs), also display multiple emission lanes. One leading theory for ZBs attributes the observed multiple lanes to radio emission at the plasma upper-hybrid frequency $\nu_{\rm uh}$ that coincides with harmonics of the electron gyrofrequency $\nu_{\mathrm{ce}}$, i.e., $\nu\approx\nu_{\rm uh}\approx (\nu_{\mathrm{pe}}^2+\nu_{\mathrm{ce}}^2)^{1/2}\approx s\nu_{\mathrm{ce}}$ \citep{
1986ApJ...307..808W,2003A&A...410.1011Z,2007SoPh..241..127K,2011ApJ...736...64C,2013SoPh..284..579Z,2018A&A...618A..60K}. However, unlike the ZBs, the frequency spacing between different emission lanes in this burst is irregular and varies in time. Moreover, although the frequency turnover times of different emission lanes $t_{\rm o}$ are very close to each other, they differ by $\sim$0.5--0.8 s (orange arrows in Figure \ref{fig-radio-spec-imaging}(F) and (H)) and does not show a systematic lag in frequency as is usually present in ZBs \citep{2007SoPh..246..431C,2007SoPh..241..127K,2013ApJ...777..159Y}. Therefore, we argue that the different emission lanes are not due to harmonics of a particular plasma wave mode. Instead, they are related to different wave packets, which are triggered by the same impulsive energy release event, propagating in magnetic flux tubes with different plasma properties.

\begin{figure*}[ht]
\includegraphics[width=\textwidth]{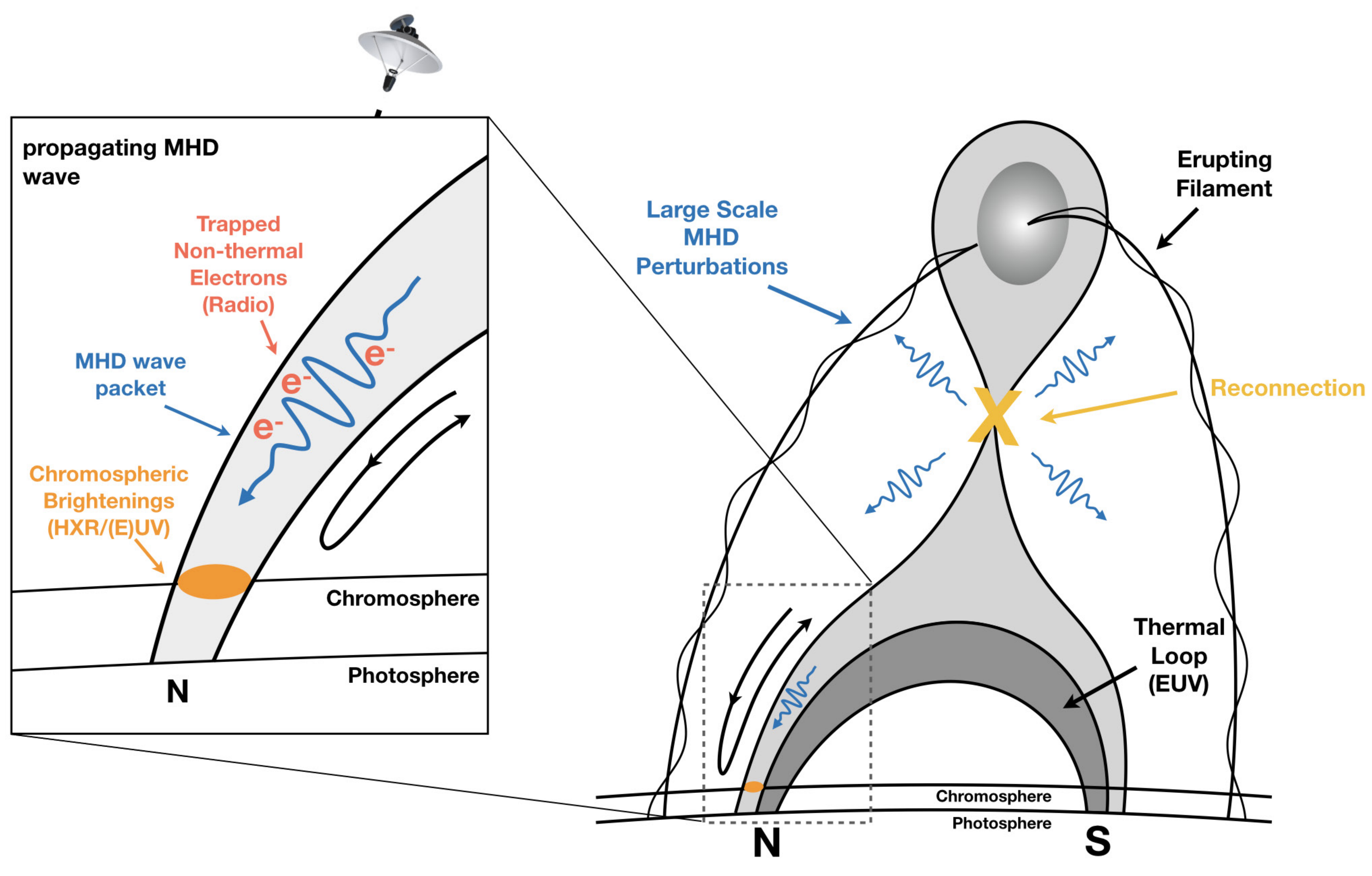}
\caption{Schematic illustration of the observed radio bursts of interest. The impulsive energy release associated with the filament eruption and the two-ribbon flare generates ubiquitous MHD disturbances, some of which propagate along newly reconnected field lines in the form of MHD wave packets that contain multiple subsecond-period oscillations. Electrons trapped or accelerated within these wave packets generate Langmuir waves and convert to radio emission. Some of the wave packets can be reflected at or near the flare ribbon due to sharp gradients, resulting in the observed spatial motion of the radio source and the low--high--low frequency drift pattern of the radio burst in the dynamic spectrum. The (E)UV brightenings at the flare ribbon may be associated with heating by the precipitated energetic electrons or the deposited wave energy. \label{fig-cartoon}}
\end{figure*}

The schematic in Figure \ref{fig-cartoon} summarizes our interpretation of the observed radio bursts in terms of propagating MHD wave packets that contain multiple subsecond-period oscillations within the context of the filament eruption and two-ribbon flare. As introduced in Section \ref{sec:intro}, subsecond-period MHD waves may \edit1{be a viable mechanism} responsible for transporting a substantial amount of the magnetic energy released in the corona downward to the lower atmosphere, resulting in intense plasma heating and/or particle acceleration. Let us consider the scenario of fast-mode MHD waves guided by dense magnetic flux tubes as an example \citep{1983SoPh...88..179E,2003A&A...409..325C}. The kinetic energy flux associated with propagating MHD waves can be estimated as $F_K\approx\frac{1}{2}\rho\delta v^2 v_g$ \citep{2007Sci...317.1192T,2014ApJ...795...18V}, where $\rho\approx m_Hn_e$ is the mass density, $\delta v$ is the amplitude of the velocity perturbation, and $v_g$ is the group speed of the propagating MHD wave. Estimates for both $\rho$ and $v_g$ can be conveniently obtained from our observations of the radio emission frequency and the radio source motion. Although the velocity perturbation $\delta v$ is not directly measured by our observations, it is intimately related to the observed density perturbation amplitude $\delta\rho \approx m_H \delta n_e$ through the continuity equation in the regime of small perturbation:
\begin{equation}
    \frac{d(\delta\rho)}{dt}=-\rho_0 \nabla \cdot \delta v,
\end{equation}
It is beyond the scope of the current study to examine the detailed relation for all possible wave modes propagating in coronal loops with different density profiles. Nevertheless, under typical coronal conditions, it has been shown by previous studies that, under typical coronal conditions, $\delta v/v_g$ is of the same order of magnitude as $\delta n_e/n_e$ for fast-mode MHD waves propagating along dense coronal loops \citep{2003A&A...409..325C, 2008ApJ...676L..73V}. The latter is found to be $\delta n_e/n_e\approx2\delta\nu/\nu\approx$\edit1{2--4$\%$}. Following these assumptions, we estimate the energy flux as \edit1{(2--8)$\times 10^8$} erg s$^{-1}$ cm$^{-2}$, with $n_e\approx 2\times 10^{10}$ cm$^{-3}$, $\delta v/v_g \approx $\edit1{2--4$\%$}, and $v_g\approx 3$ Mm s$^{-1}$.

\begin{figure*}[htb!]
\includegraphics[width=\textwidth]{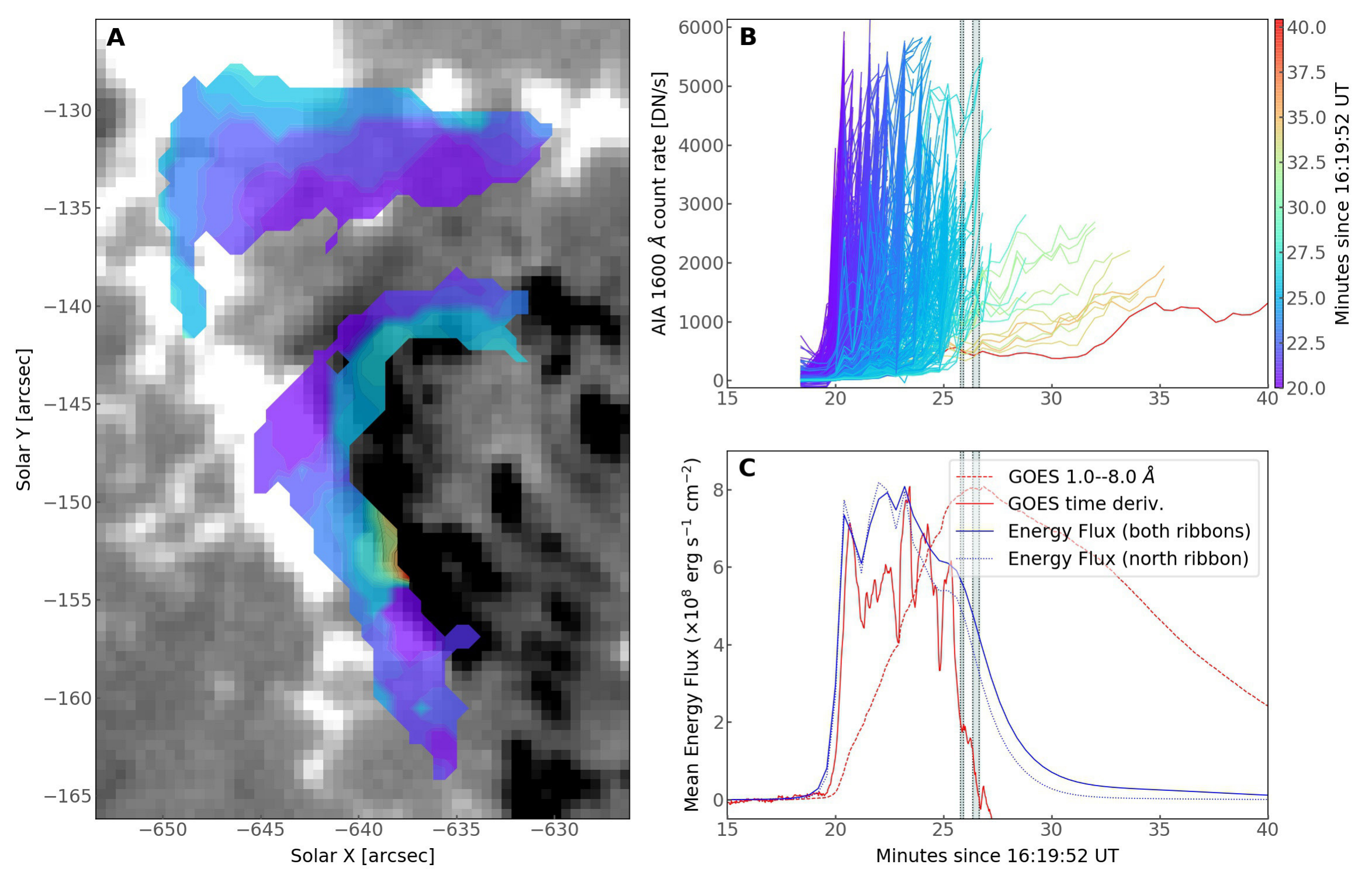}
\caption{UV observations of the flare ribbons and the associated energy flux. (A) Temporal evolution of the double UV flare ribbons in \textit{SDO}/AIA 1600 \AA, colored from purple to red in time, overlaid on the \textit{SDO}/HMI longitudinal magnetogram at 16:46:13 UT (grayscale background). (B) \textit{SDO}/AIA 1600 \AA\ light curves of all pixels on the flare ribbons (within the colored areas in (A); only the ascending part is shown). (C) \textit{GOES} 1--8 \AA\ light curve (red dashed) and its time derivative (red). \edit1{Also shown is the evolution of the energy flux of flare heating inferred from the observed UV 1600 \AA\ emission from the flare ribbon, averaged over both ribbons (blue) and the north ribbon only (blue dashed)}. The times of the two radio bursts in Figure \ref{fig-overview}(A) are demarcated with vertical lines in panels (B) and (C).\label{fig-ribbon}}
\end{figure*}

Is the estimated energy flux carried by the MHD disturbances energetically important in this flare? The energy flux required to power flares can be inferred using a variety of observational diagnostic methods including broadband imaging of flare ribbons in WL and UV \citep{2007ApJ...656.1187F,2012ApJ...752..124Q,2013ApJ...770..111L}, as well as HXR spectroscopic and imaging observations of flare footpoints \citep{2007ApJ...656.1187F}. Here we adopt the method developed by \citet{2012ApJ...752..124Q} to estimate the energy flux \edit1{needed to account for flare heating} based on \textit{SDO}/AIA 1600 \AA\ UV observations of the flare ribbons. The energy flux $F_i(t)$ of flare heating is related to UV 1600 \AA\ ribbon brightening at pixel $i$ as
\begin{equation}
    F_i(t)=\lambda I_i^{\rm pk} \exp\left[-\frac{(t-t_i^{\rm pk})^2}{2\tau_i^2}\right] \mathrm{erg\ s^{-1}\ cm^{-2}},
\end{equation}\label{eq-energy}

\noindent where the exponential term is the Gaussian function used to fit the rise phase of the light curve of the UV count rate that has a characteristic rise time $\tau_i$ and peaks at $t_i^{\rm pk}$, and $\lambda$ is the scaling factor that converts the observed peak UV count rate $I_i^{\rm pk}$ at pixel $i$ (in DN s$^{-1}$ pixel$^{-1}$, where DN is data numbers) to \edit1{the estimated energy flux responsible for the flare heating (erg s$^{-1}$ cm$^{-2}$), which depends not only on the mechanism of UV radiation upon heating in the lower atmosphere, but also on the instrument response. \citet{2012ApJ...752..124Q} and \citet{2013ApJ...770..111L} performed detailed modeling studies of loop heating of two flares, and found that $\lambda$ generally lies in the range (2-3)$\times10^5\ \mathrm{erg\ DN^{-1}\ pixel/cm^{-2}}$ to best match the model-computed \textit{GOES} SXR light curves with the observations. Here we take $\lambda \approx 2.7\times10^5\ \mathrm{erg\ DN^{-1}\ pixel/cm^{-2}}$ quoted in \citet{2012ApJ...752..124Q} for our order-of-magnitude estimate}. We have traced all pixels in AIA 1600 \AA\ UV images that show flare ribbon brightenings, which are shown in Figure \ref{fig-ribbon}(A) colored by their peak time $t_i^{\rm pk}$ from purple to red. The flare ribbons show an evident separating motion during the impulsive phase of the flare, which is characteristic of two-ribbon flares and has been considered as one of the primary evidence for magnetic-reconnection-driven flare energy release \citep{2002ApJ...565.1335Q}. The corresponding light curves of the UV count rate for all ribbon pixels are shown in Figure \ref{fig-ribbon}(B), again colored by their peak time (only the rising portion of the light curve is shown). The UV ribbon brightenings agree very well in time with the \textit{GOES} SXR derivative (thick red curve in Figure \ref{fig-ribbon}(C)), suggesting that heating of the flare loops is mainly driven by the ``evaporation'' of the heated chromospheric plasma. The estimated energy flux averaged over all ribbon pixels $\overline{F}(t)$ based on Eq. \ref{eq-energy} is shown as the blue curve in Figure \ref{fig-ribbon}(C). \edit1{Also shown is the average $\overline{F}(t)$ estimated using only pixels of the northern ribbon (dashed blue curve), with which the radio bursts appear to be associated temporally and spatially (see Figure \ref{fig-ribbonbrightenings}).} The values are in the range of $10^8$--$10^9$ $\mathrm{erg\ s^{-1}\ cm^{-2}}$, which is typical for \textit{GOES} C-class flares. At the time of the radio burst, the average $\overline{F}(t)$ \edit1{at the northern ribbon} is about \edit1{$4\times 10^8\ \mathrm{erg\ s^{-1}\ cm^{-2}}$}, which is comparable to the estimated energy flux carried by the observed subsecond-period MHD wave packets. 

We note, however, taht such coherent-burst-emitting waves can only be observed when the following conditions are met. \edit1{(1) Flare-accelerated electrons are present in the vicinity of the MHD waves. (2) Conditions are satisfied for inducing nonlinear growth of Langmuir waves and the subsequent conversion to transverse radio waves. (3) The radio waves are emitted within the bandwidth of the instrument (1--2 GHz in our case). (4) The instrument is sensitive enough to distinguish the radio bursts from the background active region and flare emission---the flux density of the bursts is only $\sim 1$ sfu (1 sfu = 10$^4$ Jansky) in our case, which is barely above the noise level of most non-imaging solar radio spectrometers.}  For these reasons, the radio bursts appear relatively rare, and thus their volume filling factor in the entire flaring region is essentially unknown. \edit1{Moreover, although possible signatures of wave damping seem to be present in some bursts that we observe (see Figure \ref{fig-wavelet}(B--E)), which may be due to energy loss during their propagation, the fraction of total energy deposited to the lower solar atmosphere from the waves remains undetermined in this study.} However, considering the presence of ubiquitous large-scale fast EUV waves throughout the active region around the same time, it is reasonable to postulate that these short-period waves are also ubiquitously present in the flaring region. If this is the case, these waves may \edit1{play a role in transporting the released flare energy during the late impulsive phase of this flare, likely alongside the accelerated electrons, and the subsequent heating of the flare ribbons and arcades}. 

\section{Conclusion}\label{sec-conclusion}
Here we report radio imaging of propagating MHD waves along post-reconnection flare loops during the late impulsive phase of a two-ribbon flare. This is based on observations of a peculiar type of \dml\ radio bursts recorded by the VLA. In the radio dynamic spectrum, the bursts show a low--high--low frequency drift pattern with a moderate frequency drift rate of $\dot{\nu}/\nu \lesssim 0.2$. VLA's unique capability of imaging with spectrometer-like temporal and spectral resolution (50 ms and 2 MHz) allows us to image the radio source at every pixel in the dynamic spectrum where the burst is present. In accordance with its low--high--low frequency drift behavior, we find that the radio source firstly moves downward toward a flare ribbon before it reaches the lowest height and turns upward. The measured speed in projection is $\sim$1--2 Mm/s, which is characteristic of Alfv\'enic or fast-mode MHD waves in the low corona. Furthermore, we find that the bursts consist of many subsecond, quasi-periodic oscillations in emission frequency, interpreted as fast oscillations within propagating MHD wave packets. As illustrated in Figure \ref{fig-cartoon}, these wave packets are likely triggered by the impulsive flare energy release, and subsequently propagate downward along the newly reconnected field lines down to the flare ribbons. From the observed density oscillations and the source motion, we estimate that these wave packets carry an energy flux of \edit1{(2--8)$\times 10^8$ erg s$^{-1}$ cm$^{-2}$}, which is comparable to the average energy flux required for driving the flare heating \edit1{during the late impulsive phase of the flare estimated from the UV ribbon brightenings. In addition, the radio source seems to show a close spatial and temporal association with the transient brightenings on the flare ribbon}. As introduced in Section \ref{sec:intro}, such subsecond-period MHD waves have long been postulated as an alternative or complementary means for transporting the bulk of energy released in flares alongside electron beams, resulting in strong plasma heating and/or particle acceleration. Here we provide, to the best of our knowledge, the first possible observational evidence for these subsecond-period MHD waves propagating in post-reconnection magnetic loops derived from imaging and spectroscopy data, and demonstrate \edit1{their possible role in driving plasma heating during the late impulsive phase of this flare event. Future studies are required to, first of all, investigate their presence in other flare events, and moreover, establish whether or not they are energetically important in transporting the released flare energy during different flare phases}. 

\acknowledgments

We thank Sophie Musset for her help in producing the \textit{RHESSI} X-ray image. We also thank Tim Bastian, John Wygant, Lindsay Glesener, Kathy Reeves, and Dale Gary for helpful discussions, \edit1{as well as an anonymous referee who provided constructive comments to improve the paper}. The National Radio Astronomy Observatory is a facility of the National Science Foundation (NSF) operated under cooperative agreement by Associated Universities, Inc. This work made use of open-source software packages including CASA \citep{2007ASPC..376..127M}, SunPy \citep{2015CS&D....8a4009S}, and Astropy \citep{2013A&A...558A..33A, 2018AJ....156..123A}. B.C. and S.Y. are supported by NASA grant NNX17AB82G and NSF grant AGS-1654382 to the New Jersey Institute of Technology.

\facilities{VLA, \textit{SDO}, \textit{RHESSI}, \textit{GOES}}
\bibliography{Yu2018}



\listofchanges

\end{document}